\documentclass[a4paper, 10pt, conference]{IEEEtran}
\IEEEoverridecommandlockouts
\usepackage[utf8]{inputenc}
\usepackage[T1]{fontenc}
\usepackage{graphicx}
\usepackage{subcaption}
\usepackage{cite}
\usepackage{amsmath,amssymb,amsfonts}
\usepackage{algorithmic}
\usepackage{graphicx}
\usepackage{textcomp}
\usepackage{xcolor}
\usepackage{float}

\usepackage{xcolor,soul}
\usepackage{xspace}
\usepackage{enumitem}
\usepackage{titlecaps}
\usepackage{tabularx}
\usepackage{balance}
\usepackage{color}
\usepackage{comment}
\usepackage{csquotes}
\usepackage{amsmath,amssymb}
\usepackage{url}
\usepackage{hyperref}
\usepackage{multirow}
\def\BibTeX{{\rm B\kern-.05em{\sc i\kern-.025em b}\kern-.08em
    T\kern-.1667em\lower.7ex\hbox{E}\kern-.125emX}}
\begin{document}

\makeatletter
\def\ps@IEEEtitlepagestyle{%
  \def\@oddfoot{\mycopyrightnotice}%
  \def\@evenfoot{}%
}
\def\mycopyrightnotice{%
  {\footnotesize \textcolor{red}{\begin{tabular}[t]{@{}l@{}} This paper has been accepted for publication by IFIP Networking 2024. The final version will appear in the conference proceedings. \\ The copyright of the final manuscript is with IFIP. \end{tabular}}}
  \gdef\mycopyrightnotice{}
}

\def\baselinestretch{0.95}

\newcommand{\TODO}[1]{\hl{\textbf{TODO:} #1}\xspace}
\newcommand{\sol}{{\em Deduplicator}\xspace}
\newcommand{\eg}{{\em e.g.,}\ }
\newcommand{\ie}{{\em i.e.,}\ }

\title{\sol: When Computation Reuse Meets Load Balancing at the Network Edge}

\vspace{-10cm}

\author{\IEEEauthorblockN{Md Washik Al Azad}
\IEEEauthorblockA{\textit{University of Notre Dame}\\
malazad@nd.edu}
\and
\IEEEauthorblockN{Spyridon Mastorakis}
\IEEEauthorblockA{\textit{University of Notre Dame}\\
mastorakis@nd.edu}
}

\makeatletter
\patchcmd{\@maketitle}
 {\addvspace{0.5\baselineskip}\egroup}
 {\addvspace{-1\baselineskip}\egroup}
 {}
 {}
\makeatother

\maketitle

\begin{abstract}

Load balancing has been a fundamental building block of cloud and, more recently, edge computing environments. At the same time, in edge computing environments, prior research has highlighted that applications operate on similar (correlated) data. Based on this observation, prior research has advocated for the direction of ``computation reuse'', where the results of previously executed computational tasks are stored at the edge and are reused (if possible) to satisfy incoming tasks with similar input data, instead of executing incoming tasks from scratch. Both load balancing and computation reuse are critical to the deployment of scalable edge computing environments, yet they are contradictory in nature. In this paper, we propose the \sol, a middlebox that aims to facilitate both load balancing and computation reuse at the edge. The \sol features mechanisms to identify and deduplicate similar tasks offloaded by user devices, collect information about the usage of edge servers' resources, manage the addition of new edge servers and the failures of existing edge servers, and ultimately balance the load imposed on edge servers. Our evaluation results demonstrate that the \sol achieves up to 20\% higher percentages of computation reuse compared to several other load balancing approaches, while also effectively balancing the distribution of tasks among edge servers at line rate.

\end{abstract}


\section {Introduction}
\label{sec:intro}

Middleboxes have been proposed and widely deployed over the years across both private and public networks~\cite{sekar2011middlebox, sherry2012making}. Among the most widely deployed middleboxes are load balancers, which are fundamental components of creating scalable web services and networked environments, such as cloud data centers~\cite{zhang2018load}. With the deployment of applications that require ultra low user-perceived latency and produce massive volumes of data (\eg augmented and virtual reality applications, smart city applications), edge computing has emerged as a paradigm that brings computing resources (small-scale data centers) physically close to user devices and applications~\cite{shi2016edge}. 

In edge computing environments, prior studies have highlighted that applications operate on temporally, spatially, and contextually similar data~\cite{guo2018potluck, guo2018foggycache, meng2020coterie}. In other words, applications at the edge may request the processing of similar data by offloading computational tasks with similar input data for the same services (\eg image or video annotation). The execution of such tasks often yields the same output/outcome (execution results), resulting in the execution of duplicate (redundant) computation. To eliminate duplicate computation and reduce task completion times, the direction of ``computation reuse'' has been proposed, where edge servers store the results of executed tasks to reuse them and satisfy incoming similar tasks (tasks with similar input data for the same services)~\cite{guo2018potluck, al2022promise}. 

In line with these observations, in this paper, we argue that both load balancing and computation reuse are vital functions for the realization of scalable edge computing environments. Nevertheless, these functions are contradictory in nature. On the one hand, load balancing aims to distribute the load equally among the available edge servers. On the other hand, computation reuse aims to distribute tasks with similar input data to the same edge server(s), which causes load imbalances, since: (i) tasks with similar input data are not uniformly generated at the edge (\eg certain applications/services may be more popular than others)~\cite{wei2021popularity, nicolaescu2021store}; and (ii) due to (i), certain edge servers may receive and execute more tasks than others.

In this paper, we shed light into the following \textbf{research questions}: ``is it possible to achieve both the, yet contradictory in nature, functions of load balancing and computation reuse in edge computing environments? If so, to what extent and what are the tradeoffs between these functions?''. To answer these questions, we propose the \sol, a middlebox that aims to facilitate the reuse of computation, while at the same time balancing the load imposed on edge servers. Our work makes the following contributions:

\begin{itemize}[leftmargin=0cm,itemindent=.3cm,labelwidth=\itemindent,labelsep=0cm,align=left, noitemsep, topsep=0pt]


\item We present the \sol middlebox design, which takes advantage of Locality Sensitive Hashing (LSH)~\cite{indyk1998approximate} to identify and deduplicate similar computational tasks offloaded by user devices. It also features mechanisms to collect information about the usage of edge servers' resources, manage the addition of new edge servers and the failures of existing ones, and ultimately balance the load imposed on edge servers.

\item We implement a \sol prototype on top of Ngnix~\cite{reese2008nginx}, a popular open-source web server, HTTP proxy, and load balancing framework~\cite{reese2008nginx}. Our experimental evaluation demonstrates that the \sol prototype identifies and deduplicates similar offloaded tasks with minor performance overhead, being able to distribute them to available edge servers at line rate. In addition, it is able to effectively balance the tradeoffs between load balancing and computation reuse at the edge, facilitating the realization of both functions.

\end{itemize}


\section{Use Cases and Motivation}
\label{sec:use-cases}



\subsection{Computation Reuse Use Cases}

\begin{figure*}[t]
\vspace{0.2cm}
	\centering
	\begin{subfigure}{.32\textwidth}
		\centering
		\includegraphics[scale=0.21]{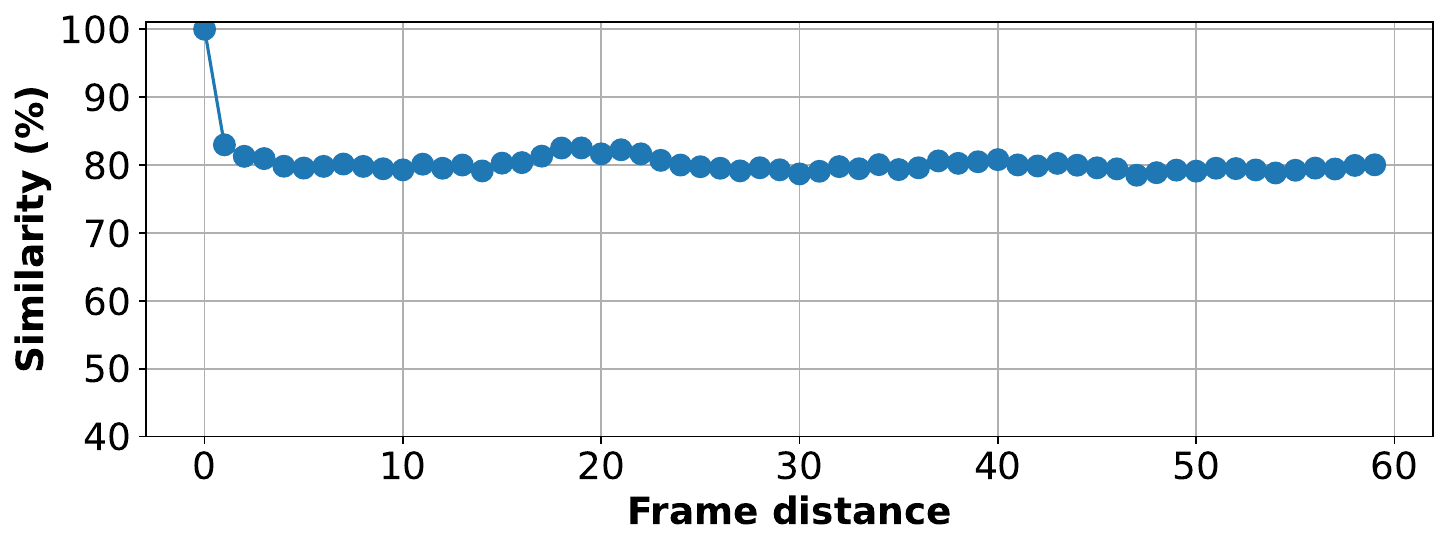}
		\vspace{-0.2cm}
		\caption{Similarity between subsequent frames when a user plays an AR game without significantly rotating their headset's camera.\\}
		\label{Figure:pokemon_go_frame_similarity}
	\end{subfigure} \hfill
	\begin{subfigure}{.32\textwidth}
		\centering
		\includegraphics[scale=0.21]{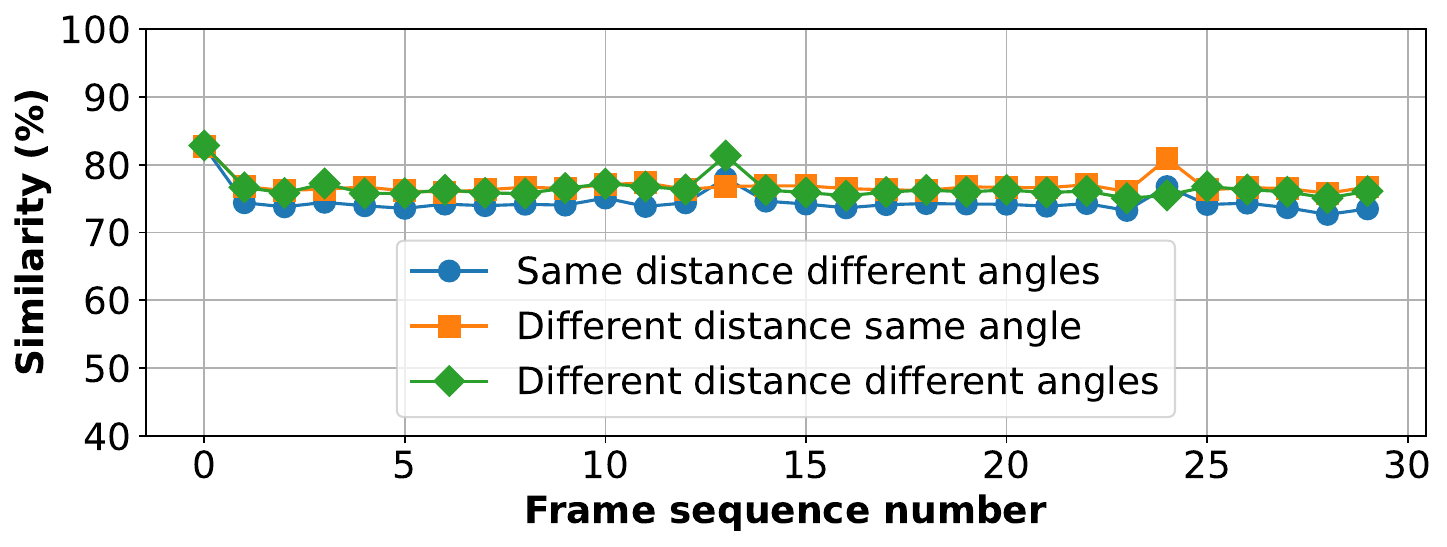}
		\vspace{-0.2cm}
		\caption{Similarity between video sequences of two players when a single object is depicted in captured frames.\\}
		\label{Figure:motivation2_single_object}
	\end{subfigure} \hfill
	\begin{subfigure}{.32\textwidth}
		\centering
		\includegraphics[scale=0.21]{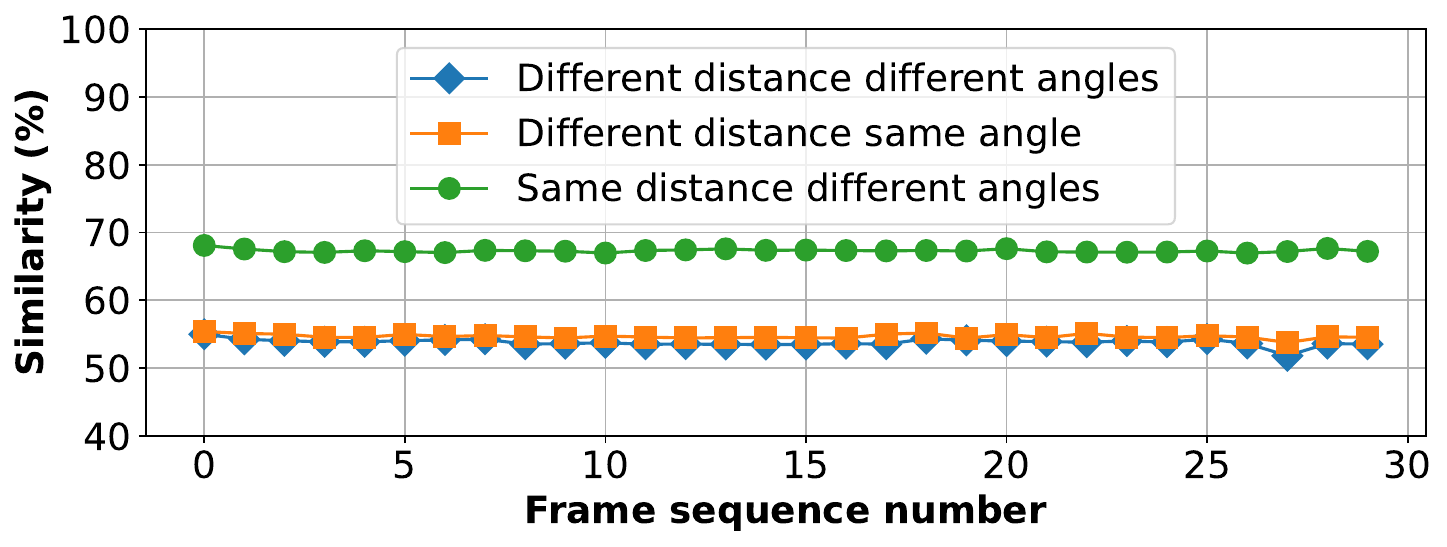}
		\vspace{-0.2cm}
		\caption{Similarity between video sequences from two players when multiple objects are depicted in captured frames.\\}
		\label{Figure:motivation2_multiple_objects}
	\end{subfigure}
	\vspace{-0.5cm}
	\caption{Similarity among video frames in different application scenarios.}
	\vspace{-0.6cm}
\end{figure*}

\noindent\textbf{Multi-player Augmented Reality (AR) gaming:} Let us consider a scenario where users play an AR game through their AR headsets. 
For virtual objects to reside into the real world, compute-heavy processing tasks, such as detecting objects in captured camera frames as well as measuring the height of objects and their distances from the headset, need to be performed. Due to the limited computing capacity of AR headsets, captured snapshots may be offloaded to an edge server, which executes compute-heavy tasks on behalf of headsets. For a satisfactory user experience, at least 30 Frames Per Second (FPS) need to be processed. With such a high FPS rate, chances are that subsequent frames will be similar to each other and depict the same scene from slightly different angles or distances. Figure \ref{Figure:pokemon_go_frame_similarity} shows that the similarity between subsequent frames is around 80\% when a user plays an AR game without significantly rotating their headset's camera. Since subsequent frames are substantially similar, they may yield the same results (outputs) when processed by an edge server (\eg to identify objects in these frames). In other words, similar enough input data could yield the same processing results (output). To this end, edge servers: (i) store the results of processed frames; and (ii) identify whether incoming frames are similar enough to previously processed ones and, if so, reuse the results stored for previously processed frames. This alleviates the need to process all frames from scratch, since incoming frames are processed from scratch only when similar enough frames have not been previously processed.


\noindent\textbf{Cognitive assistance:} Let us consider a cognitive assistance application, such as Google Lens, which allows users to search for related content by identifying the object(s) or text in a picture captured by a mobile phone. For instance, millions of people visit the Statue of Liberty each year and take pictures of this sight with their phones from different angles or distances. Visitors can use such an application to learn more about the Statue of Liberty, 
such as the history behind the statue, and related videos and podcasts. Captured pictures will be offloaded to an edge server, which will identify the depicted sight and return related search results to users. Since pictures related to a certain sight depict the same sight from different distances/angles, 
the search results will be the same.

\noindent\textbf{Traffic monitoring in a smart city:} Let us consider a smart city where CCTV cameras are deployed for traffic management purposes. Snapshots captured by these cameras are offloaded to edge servers, which run computer vision algorithms to identify the number of vehicles in each snapshot. A CCTV camera captures multiple snapshots per second, thus subsequent snapshots are similar to each other.  
Instead of blindly identifying the number of vehicles in each snapshot, an edge server identifies whether an incoming snapshot is similar to previous snapshots. If no similar previously processed snapshots are found, the edge server processes the snapshot from scratch and stores the results for future reuse. If a similar snapshot has been processed, its processing results are reused.

A commonality among the above use cases is the existence of a massive number of devices, such as smartphones, AR headsets, IoT devices (potentially in close proximity to each other), which offload computational tasks to servers at the edge of the network. For example, cities around the world, such as Beijing, London, and New York, have several hundred thousands 
of CCTV cameras, 
while every day tens of thousands of people visit popular sights around the world, such as the Statue of Liberty. 
Another commonality is that the offloaded data to be processed may be semantically, temporally, and/or spatially similar, thus the processing of this data may yield the same output, resulting in the execution of redundant computation. To this end, edge servers store the results of executed tasks, so that their results can be reused to satisfy incoming tasks with similar input data without the need to be executed from scratch. In other words, computation reuse has the goal of enabling edge computing environments to reduce (if not eliminate) the execution of redundant computation.

\vspace{-0.2cm}
\subsection{\sol Motivation}

Let us consider the multi-player AR gaming use case discussed above to showcase the necessity of \sol as a middlebox to facilitate both computation reuse and load balancing in edge computing environments. As we illustrate in Figure \ref{Figure:overview}, user 1 offloads two similar subsequent frames/snapshots (1a and 1b) captured by an AR headset to edge servers, and these frames are distributed to two different servers. In this case, it is not possible to reuse the results of the first frame's processing to satisfy the processing of the second frame, thus both frames need to be processed from scratch. To achieve computation reuse, we need mechanisms to introduce reuse awareness in the task distribution process, so that tasks with similar inputs are distributed to the same edge server(s). 




Let us further consider the case of users who play the AR game in an collaborative environment (\eg users 1, 2, and 3 in Figure~\ref{Figure:overview}). Users can meet at a physical location in the real world, where they interact with virtual objects in a collaborative manner (\eg Pokemon Go application). The frames offloaded by these players are similar to each other as presented in Figures \ref{Figure:motivation2_single_object} and \ref{Figure:motivation2_multiple_objects}, since they depict the same physical location from different angles/distances. As such, the corresponding tasks should be distributed to the same edge server to facilitate computation reuse (\eg tasks 1a, 2, and 3 in Figure~\ref{Figure:overview}). Distributing similar tasks to the same edge server(s) for a wide spectrum of applications may overload certain servers, while leaving others underutilized, thus causing load imbalances among servers. 
This signifies the need in edge computing environments to facilitate computation reuse, while at the same time balancing the load imposed on edge servers. To this end, in this paper, we propose the \sol, a middlebox that aims to simultaneously achieve these functions. 




\noindent \textbf{How is the \sol different than traditional load balancing:} 
Several load balancing techniques have been proposed over the years. These techniques do not consider the similarity between the input data of tasks, thus they do not facilitate computation reuse during the distribution of tasks towards servers. For example, load balancing based on consistent hashing~\cite{wang2007load} provides a session-persistent connection between a user and a server by hashing the header fields of user requests. A naive solution could be to replace the hash function used (consistent hashing) with LSH and hash the task input data to forward tasks with similar inputs to the same server(s). Although this approach increases the probability of reuse, as our evaluation results show in Section \ref{subsec:results}, it will also cause severe load imbalances among servers. On the other hand, the \sol provides mechanisms to simultaneously facilitate computation reuse and balance the load among edge servers.

\section{\sol Design}
\label{sec:design}


\subsection{Design Overview}
\label{subsec:overview}

We assume the existence of small-scale data centers at the edge, called ``cloudlets''~\cite{satyanarayanan2009case}. A cloudlet consists of heterogeneous servers with computing and storage resources. Each server may offer a set of services (\eg object recognition, face detection) to users or execute the code of processing functions offloaded directly by user devices.

\begin{figure}[!t]
\vspace{0.07cm}
 \centering
 \includegraphics[width=0.7\columnwidth]{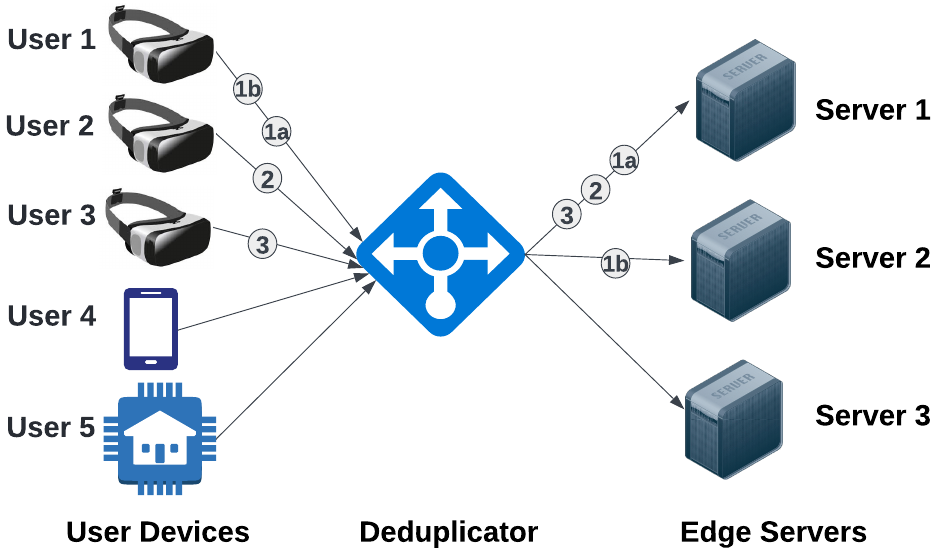}
 \vspace{-0.2cm}
 \caption{An edge computing environment where the \sol receives tasks offloaded by user devices and distributes them to edge servers.}
 \label{Figure:overview}
 \vspace{-0.6cm}
\end{figure}

The \sol operates as a middlebox in edge computing environments, which is placed in front of available
edge servers (Figure~\ref{Figure:overview}). It aims to achieve the following \textbf{functions}: \textbf{(i) balance the load among available edge servers}; and \textbf{(ii) facilitate the reuse of computation}. In the context of (i), the \sol distributes computation tasks among the available
edge servers in ways that equalize the load among them. In the context of (ii), the \sol distributes similar tasks to the same edge servers, enabling servers to reuse the results of tasks that they have previously executed. 

User devices offload computation tasks as HTTP/HTTPS POST requests, which identify the requested services (in the URL field) and carry the input data (in the request body). In the case of HTTPs, a user device first creates a TLS session with the middlebox so that the requested service and input data can be extracted. The \sol intercepts HTTP/HTTPS POST requests and makes use of Locality Sensitive Hashing (LSH)~\cite{indyk1998approximate} to identify requests for similar computation, which can be reused by edge servers. To achieve that, the \sol slices the space of LSH values and distributes (assigns) these slices among edge servers. Each edge server becomes responsible for executing tasks with input data hashed to values that fall in the slice assigned to it. The \sol collects information about the usage of server resources over time and redistributes the hash value space among edge servers, so that the load is balanced while facilitating computation reuse. Finally, it features mechanisms to identify newly added servers and server failures, and redistributes the hash value space to achieve its primary functions.

\subsection{Task Offloading and Execution Process}
\label{subsec:offloading}

Tasks are created by user devices and are offloaded to edge servers. Each task is represented as an HTTP/HTTPS POST request. In the URL field, the URL of the requested service is included. The input data to be processed is included in the request's body. 
Each task also carries a similarity threshold, so that applications can indicate the minimum threshold of similarity with tasks previously executed by edge servers, which would be acceptable for reuse. We discuss how an appropriate similarity threshold can be determined for an application below. An offloaded task will be received by the \sol, which is placed in front of available edge servers and distributes the task towards one of these servers based on the locality sensitive hash of the task's input data. This hash can be produced based on one of the following mechanisms:

\begin{itemize}[leftmargin=0cm,itemindent=.3cm,labelwidth=\itemindent,labelsep=0cm,align=left, noitemsep, topsep=0pt]

\item {\textbf{Hashing by the \sol:} } The \sol receives an offloaded task and hashes its input data through LSH. This mechanism is fully transparent to users, since LSH is performed by the \sol. However, this could incur overhead on the \sol, which aims to distribute offloaded tasks at line rates.

\item {\textbf{User assisted hashing:}} Users apply LSH to the task input data and attach the hash value to the task. This mechanism requires user involvement, however, it does not require the \sol to apply LSH to the task input data.

As we mentioned in Section~\ref{subsec:overview}, the \sol slices the space of potential LSH values and assigns these slices among edge servers. In other words, each edge server becomes responsible for executing tasks with input data hashed to values that fall in the slice assigned to it. For example, as we explain in Section~\ref{subsec:slicing} and illustrate in Figure~\ref{Figure:scenario1}, the first edge server $S_1$ is responsible for executing tasks with input data hashed to values 0 - 21844. Once the \sol receives a task, it will produce the locality sensitive hash of the task's input data (or extract the hash attached to the task) and distribute the task based on the resulting hash value. For example, if the hash value of a task's input data falls under the slice of $S_1$, the task will be distributed to $S_1$.

Once an edge server receives an offloaded task, it will search for previously executed similar tasks that have been stored by the server. If such a task is found, its execution results are returned to the user device that offloaded the task. For the selection of a previously executed similar task, edge servers will make use of the similarity threshold attached to tasks. Specifically, in response to a new offloaded task $t_{new}$, edge servers will select and return a stored task $t_{stored}$ for the same service as $t_{new}$, so that: (i) the input data of $t_{stored}$ has the highest similarity (nearest neighbor) among stored tasks to the input data of $t_{new}$; and (ii) the similarity between the input data of $t_{stored}$ and $t_{new}$ exceeds the similarity threshold that is acceptable by the application that offloaded $t_{new}$. If there is no previously executed task that satisfies these conditions, $t_{new}$ will be executed from scratch and its execution results will be stored by the server for reuse in the future.

\noindent\textbf{Deciding on appropriate similarity thresholds:} For a specific application, a developer or a service provider can select an approximate initial similarity threshold value based on the application's nature. Edge servers will initially reuse computation based on this approximate threshold, and the \sol will adjust the threshold by sampling the reuse accuracy of groups of tasks over time. The sampling process can take place by having servers execute from scratch small groups of tasks $g_1, g_2, ..., g_n$ for which they have results that they could potentially reuse and compare the results of the execution of $g_1, g_2, ..., g_n$ with the results of the tasks that could be reused. Based on this process, the \sol can estimate the reuse accuracy and adjust the similarity threshold accordingly. 

\end{itemize}


\subsection{Hash Value Space Slicing}
\label{subsec:slicing}

For every hash function, there is a space (range) of resulting hash values. The same applies to the \sol, which slices the LSH value space and distributes slices to available edge servers. Each edge server receives and executes offloaded tasks with input data resulting in hash values that fall in the slice assigned to this server when hashed through LSH. 

\begin{figure*}[!t]
    \captionsetup[subfigure]{aboveskip=-0.1pt,belowskip=-0.1pt}
	\centering
	\begin{subfigure}{0.21\textwidth}
		\centering
		\includegraphics[scale=0.27]{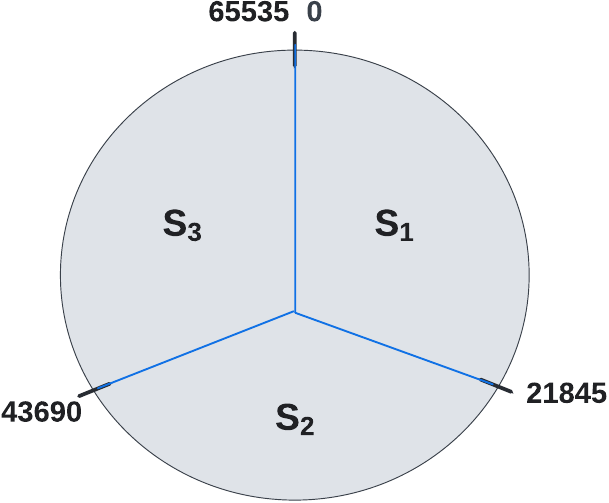}
		\vspace{0.01cm}
		\caption{A scenario of an equal distribution of the hash value space among edge servers.} \hfill
		\label{Figure:scenario1}
	\end{subfigure} 
	\hfill
	\begin{subfigure}{0.21\textwidth}
		\centering
		\includegraphics[scale=0.27]{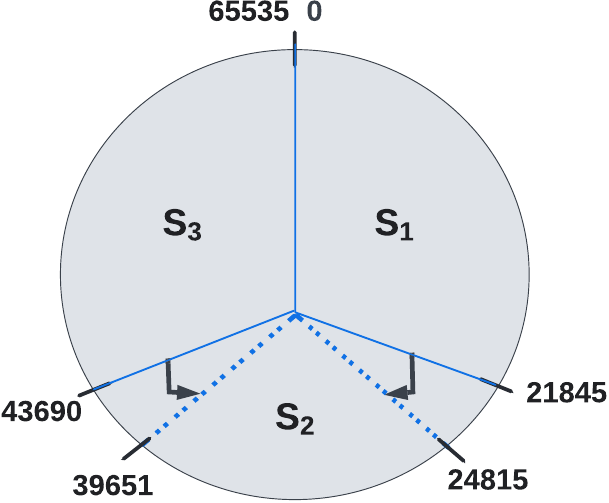}
		\caption{A scenario where $S_2$ becomes overloaded and the slice assigned to $S_2$ shrinks.} \hfill
		\label{Figure:scenario2}
	\end{subfigure} 
	\hfill
	\begin{subfigure}{0.22\textwidth}
		\centering
		\includegraphics[scale=0.27]{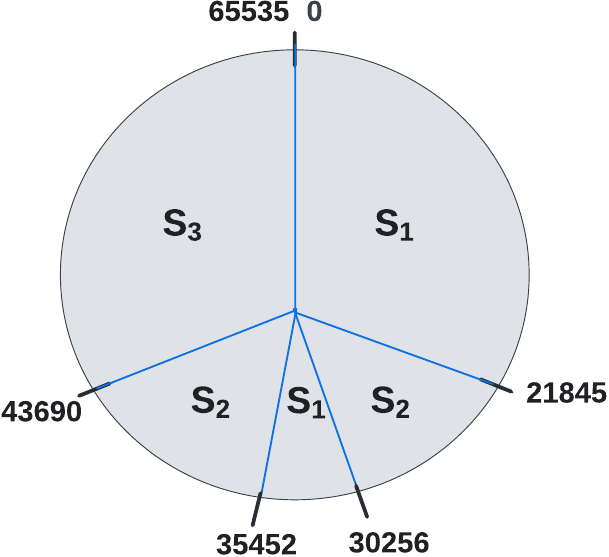}
		\vspace{0.01cm}
		\caption{A scenario where $S_2$ becomes overloaded and slices of finer granularity are created.} \hfill
		\label{Figure:scenario3}
	\end{subfigure}
	 \hfill
	 \begin{subfigure}{0.26\textwidth}
	 	\centering
	 	\includegraphics[scale=0.25]{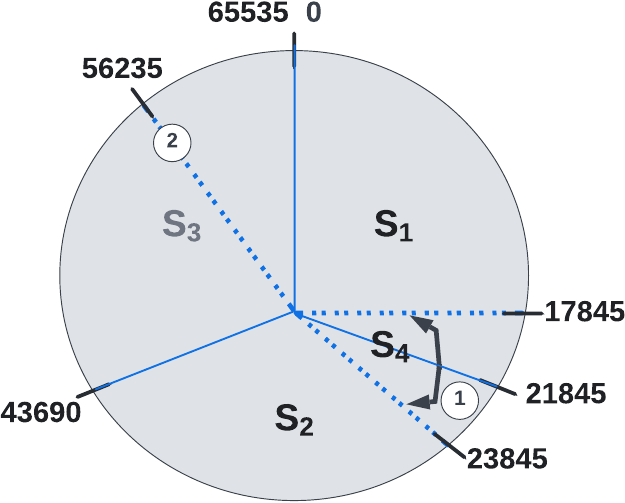}
            \vspace{0.3cm}
	 	\caption{A scenario where the hash space is redistributed due to (1) the addition and (2) the failure of a 
   server.} \hfill
         \label{Figure:scenario4}
	 \end{subfigure}
	\vspace{-0.45cm}
	\caption{Distribution (slicing) of a hash value space (hash length equal to 2 bytes) among three edge servers.}
	\label{Figure:distribution}
	\vspace{-0.6cm}
\end{figure*}

In Figure~\ref{Figure:scenario1}, we present an example of a hash space with a hash length equal to 2 bytes (potential values 0 - 65535). In this example, the \sol has distributed the space equally among the three available edge servers: the first slice includes hash values 0 - 21844 and is assigned to the first server ($S_1$), the second slice includes hash values 21845 - 43689 and is assigned to the second server ($S_2$), and the third slice includes hash values 43690 - 65535 and is assigned to the third server ($S_3$). 
The \sol forwards tasks to servers according to input data hash values: 0 - 21844 to Server $S1$, 21845 - 43689 to Server $S2$, and 43690 - 65535 to Server $S3$.

The \sol begins its operation with an equal distribution of the hash space among 
edge servers. Such an equal distribution may cause load imbalances among edge servers, since the distribution of tasks in the hash space is unlikely to be uniform over time. 
Over time, the \sol collects usage information about the resources of edge servers (\eg CPU and memory usage), identifies overloaded and underutilized edge servers, and redistributes the hash value space among servers 
to balance the load without impairing the ability of the edge computing environment to achieve computation reuse. 

\noindent \textbf{Hash value space redistribution strategies:} A challenge that arises during the redistribution process of the hash value space has to do with which values in a slice (and the associated tasks) impose substantial load on each edge server. To address this challenge, the \sol employs two strategies to redistribute the hash value space among edge servers in order to alleviate load imbalances: 

\begin{itemize}[leftmargin=0cm,itemindent=.3cm,labelwidth=\itemindent,labelsep=0cm,align=left, noitemsep, topsep=0pt]

\item \textbf{Adjusting the size of existing slices:} This strategy applies to cases where significant load is imposed due to hash values (and associated tasks) close to one or both ``edges'' of a slice. For example, in Figure~\ref{Figure:scenario2}, we present a scenario where $S_2$ becomes overloaded and a substantial part of the load is due to values close to the edges of the slice assigned to $S_2$. In this case, one or both edges of the slice can be redistributed to edge server(s) possessing adjacent slices ($S_1$ and/or $S_3$ in our example). In other words, the slice assigned to the overloaded server becomes smaller, while the sizes of adjacent slices grow.

\item \textbf{Creating slices of finer granularity:} This strategy applies to cases where significant load is imposed due to hash values (and associated tasks) that are not close to an edge of a slice. For example, in Figure~\ref{Figure:scenario3}, we present a scenario where $S_2$ becomes overloaded and a substantial part of the load is due to values closer to the middle of the slice assigned to $S_2$. In this case, the \sol creates slices of finer granularity. Specifically, it creates a slice with values in the middle of the initial slice assigned to $S_2$. As a result, the newly created slice (which includes hash values 30256 - 35452 in our example) will be assigned to another server ($S_1$ in our example). In this way, the offloaded tasks that fall under this new slice (and subsequently part of the load imposed to $S_2$) will now be distributed by the \sol to $S_1$. The \sol can also perform the reverse process, thus being able to aggregate multiple fine-grained slices to create a coarse-grained slice, if needed during its operation.

\end{itemize}

Both strategies aim to redistribute part of the load from overloaded edge servers to servers with available resources. This is achieved by adjusting the created slices, so that part(s) of the hash value space are re-assigned (from overloaded servers to servers with available resources).  As a result, tasks are distributed by the \sol to 
edge servers based on the updated slices. 

When a redistribution of the hash value space takes place, offloaded tasks with input data resulting in the redistributed hash values will be executed by a different edge server. For example, offloaded tasks will be executed by $S_1$ instead of $S_2$ for the redistributed hash values in the scenarios of Figures~\ref{Figure:scenario2} and~\ref{Figure:scenario3}. In such cases, the execution results of tasks, which correspond to the redistributed hash values and are stored by $S_2$ in order to be reused, can be transferred from $S_2$ to $S_1$. This enables $S_1$ to reuse the results of previously executed tasks that correspond to redistributed hash values right away (warm start) instead of having to first execute offloaded tasks from scratch for the redistributed hash values in order to store their results and reuse them in the future (cold start). 

\subsection {Collecting Resource Usage Information From Edge Servers}

To redistribute the hash value space, the \sol needs to collect information about the resource usage of available edge servers over time. Such information may include CPU and memory usage, among others. The collection of this information is achieved through the following mechanisms: 

\begin{itemize}[leftmargin=0cm,itemindent=.3cm,labelwidth=\itemindent,labelsep=0cm,align=left, noitemsep, topsep=0pt]

\item {\textbf{Explicit notifications:}} Edge servers send explicit notifications to the \sol over time to indicate their current resource usage. This can happen either periodically or once the resource usage of an edge server exceeds a certain threshold.

\item {\textbf{Piggybacking information on HTTP/HTTPS POST responses:}} Edge servers piggyback resource usage information on HTTP/HTTPS POST responses, which contain the results of the execution of offloaded tasks. Once a response that carries such information is received by the \sol, the \sol extracts this information from the response and the response is returned to the user device that offloaded the corresponding task (HTTP/HTTPS POST request). 

\end{itemize}

Ideally, we would like the \sol to collect CPU and memory usage information per hash value. However, as the size of the hash value space increases, it becomes infeasible to maintain such information on a per hash value basis. To this end, edge servers maintain such information for a range (group) of hash values. The size of this range is typically smaller than an individual slice (finer granularity) and it is determined, so that it provides sufficient granularity without imposing significant overhead. 

\subsection {Edge Server Additions and Failures}

The \sol is able to identify additions of new edge servers and failures of existing servers. In both cases, the hash value space needs to be redistributed. When adding a new server, a slice of the hash value space needs to be created and assigned to the newly added server. An approach to achieve that is for the \sol to create a new slice out of the slice of the most loaded edge server. For example, in Figure~\ref{Figure:scenario4}, let us assume that the most loaded edge server is $S_2$ and a new edge server $S_4$ is added. The \sol will re-assign a part of $S_2$'s slice to $S_4$, thus creating a slice for $S_4$. Optionally, the \sol can also re-assign a part of a slice adjacent to $S_2$'s slice (the slice of $S_1$ in our example) to $S_4$ in order to create a slice of a more substantial size, so that additional load is handled by $S_4$ (more offloaded tasks are distributed to $S_4$). The exact part of the hash value space to be assigned to $S_4$ depends on the load imposed by the different hash values in the slice of $S_2$, and potentially the different hash values in the slice adjacent to $S_2$'s slice.

In the case of an edge server failure ($S_3$ in the example of Figure~\ref{Figure:scenario4}), the \sol will re-assign parts of the slice of the failed server to adjacent slices. In our example, the \sol will re-assign a part of $S_3$'s slice to $S_1$ and the remaining part of $S_3$'s slice to $S_2$. The \sol aims to re-assign the slice of $S_3$ to adjacent slices, so that the corresponding servers ($S_1$ and $S_2$) do not become overloaded (based on their current load and the load to be re-assigned). Load imbalances that occur after the initial redistribution of the hash value space described above (due to the addition of a new server or the failure of an existing server) will be alleviated through subsequent redistribution events of the hash value space among servers as described in Section~\ref{subsec:slicing}. 


During the addition of a new edge server, the new server sends an explicit notification to the \sol. Once this notification is received by the \sol, the process of creating a slice for the new server begins. On the other hand, server failures are identified due to the absence of: (i) responses to offloaded tasks distributed by the \sol; and (ii) explicit notifications sent by the edge servers to the \sol to indicate the usage of their resources.

\section {Evaluation}
\label{sec:eval}


\subsection {\sol Prototype Implementation}
\label{subsec:prototype}

\begin{figure}[!t]
  \centering
  \includegraphics[width=0.90\columnwidth]{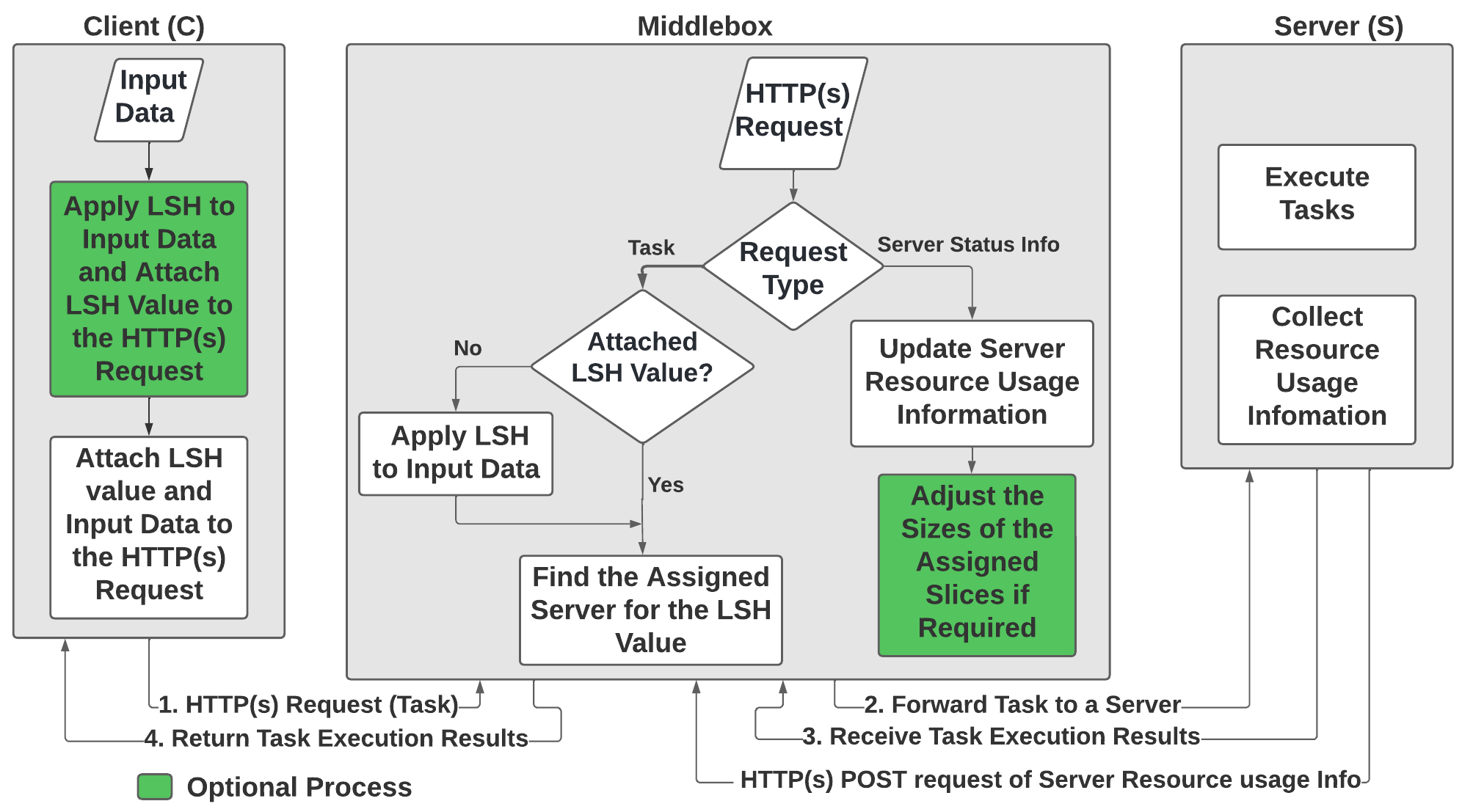}
  \vspace{-0.2cm}
  \caption{Overview of the \sol implementation.} 
  \label{Figure:system_overview}
 \vspace{-0.5cm}
\end{figure}

We implemented a \sol prototype \footnote{We make our \sol implementation code available to the research community at \url{https://github.com/malazad/Deduplicator}.}
as a module of Nginx
~\cite{reese2008nginx}. The \sol is deployed as a middlebox. 
Figure \ref{Figure:system_overview} shows an overview of our \sol implementation and how the \sol interacts with clients and servers. The \sol mainly performs two operations: (i) receives and forwards HTTP/HTTPS requests (tasks) from users to edge servers based on the assigned hash value ranges (slices) and resource availability of the servers; and (ii) collects resource usage information from servers and updates the sizes of assigned slices as described in Section \ref{sec:design}. 

The task distribution process begins when the middlebox receives HTTP/HTTPS requests from users. Depending on whether users hash the task input data, the \sol either hashes the input data attached to the received HTTP/HTTPS requests or extracts the hash values of input data from the received requests to decide to which server a task should be distributed in order to facilitate reuse. 

To redistribute slices of the hash value space, servers send resource utilization information to the \sol. 
Subsequently, the \sol updates the resource information of servers and decides whether redistribution is needed. If redistribution is needed, the slice sizes are adjusted and the stored execution results of tasks that correspond to the adjusted hash space may be transferred to a new server from previously assigned server(s) as described in Section \ref{subsec:slicing}. 


We have implemented an adaptive approach, so that our \sol prototype can adjust the granularity of slices. Let us assume that there are $n$ edge servers available at a certain time and the LSH hash size is $b$ bits. Initially, the assigned hash space per edge server will be $\frac{2^b}{n}$. If an edge server $S_i$, where $1 \leq i \leq n$, receives $t_i$ tasks during the interval between two consecutive hash space redistribution events, then the size of the hash space $H_i$ assigned to $S_i$ will be: \vspace{-0.2cm}\begin{equation} \label{eq1}
\begin{split}
H_i & = \frac{2^b}{n}  + 2^b (\frac{1}{n} - \frac{t_i}{\sum_{i=1}^{n} t_i} ) \\
\end{split}
\vspace{-0.5cm}
\end{equation}

The hash value range $R_i$ assigned to $S_i$ will be: \vspace{-0.2cm}\begin{equation} \label{eq2}
R_i =
\begin{cases}
\textbf{[}max(R_{i-1}) + 1,  max(R_{i-1})+ H_i \textbf{]}, & \text{if } 1 < i \leq n \\
\textbf{[}0, H_i -1\textbf{]}, & \text{if } i = 1
\end{cases}
\end{equation}

The second part of Equation~\ref{eq1} determines whether the range $R_i$ assigned $S_i$ needs to be adjusted, so that the slices assigned to edge servers become more fine- or coarse-grain. 

We implemented a user traffic generator, which attaches input data 
to HTTP/HTTPS POST requests (tasks) and offloads certain numbers of tasks per second. We implemented the LSH semantics through the FALCONN library \cite{andoni2015practical}. 
We implemented edge server programs, which perform different data processing operations (\eg object detection)
and collect information about the resource usage of edge servers.

\subsection{Evaluation Setup}

\begin{figure}[!t]
  \centering
  \includegraphics[scale=0.3]{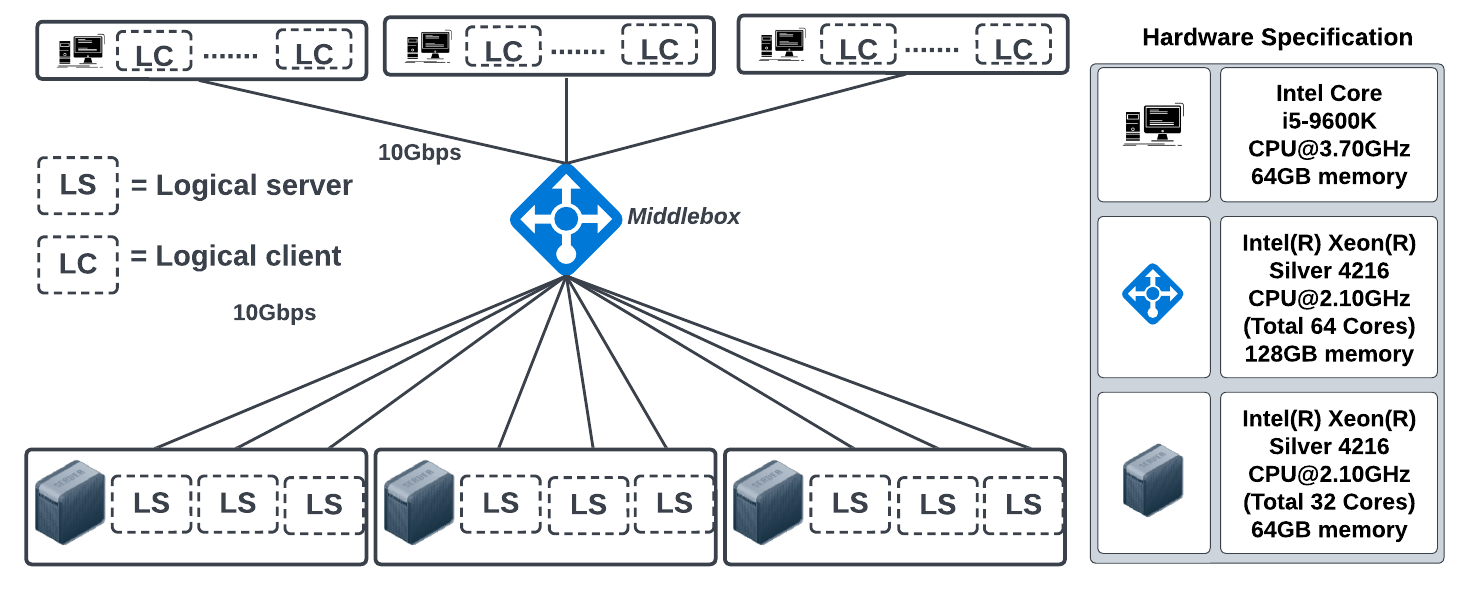}
   \vspace{-0.3cm}
  \caption{Experimental topology.} 
  \label{Figure:topology}
\end{figure}

\noindent\textbf{Experimental environment:} We deployed the \sol prototype on the testbed of Figure \ref{Figure:topology} to evaluate its performance and tradeoffs. We used three physical machines as traffic generators to emulate users who offload tasks. Each generator has hundreds of logical clients, and each logical client sends a specific number of requests (tasks) per second. We also used three physical servers, which offer different services, such as object detection and recognition. 
Each physical server hosts three logical servers to scale up our experimental setup. All traffic generators and servers are connected to the middlebox through 10Gbps Ethernet links. We run each experiment ten times and we present the average results in Section~\ref{subsec:results}.

\noindent\textbf{Datasets:} To evaluate the \sol prototype, we use five real-world image datasets as task input data with different degrees of correlation between data (Table \ref{tab:datasets}). Each edge service 
may also require a different processing granularity. 

\begin{table}[!t]
\centering
\caption{Edge services and real-world datasets used for the evaluation of the \sol.}
\label{tab:datasets}
\resizebox{1\columnwidth}{!}{%
\centering

\begin{tabular}{|c|c|c|c|c|}
\hline

\textbf{Dataset}                                              &   \textbf{\begin{tabular}[c]{@{}c@{}}Dataset size\end{tabular}} & \textbf{\begin{tabular}[c]{@{}c@{}}Correlation between data\end{tabular}} & \textbf{\begin{tabular}[c]{@{}c@{}} Edge services\end{tabular}}                                                                        & \textbf{\begin{tabular}[c]{@{}c@{}}Processing granularity\end{tabular}} \\ \hline
MNIST~\cite{lecun1998mnist}                                                         &                                                                                                       70K                                                              & Low                                                                          & Digit recognition                                                                            & Medium                                                             \\ \hline
ASL~\cite{asl}                                                                                      & 87K                                                              & Low                                                                          & \begin{tabular}[c]{@{}c@{}}Recognition  of ASL letters \end{tabular} & Medium                                                               \\ \hline
\begin{tabular}[c]{@{}c@{}}Hololens 
~\cite{hololens}\end{tabular} &                                                                                                                  189K                                                               & High                                                                     & Action recognition                                                                          & Fine                                                             \\ \hline

Pandaset~\cite{pandaset}                                                                                        & 49K                                                              & Low                                                                          & \begin{tabular}[c]{@{}c@{}}Obstacle detection  \end{tabular} & Fine                                    \\ \hline


CCTV                                                                                       & 10K                                                              & High                                                                          & \begin{tabular}[c]{@{}c@{}}Vehicle detection   \end{tabular} & Coarse                                     \\ \hline


\end{tabular}
}
\vspace{-0.3cm}
\end{table}

\noindent\textbf{Evaluation metrics:} 
We consider the following metrics:


\begin{enumerate} [wide, labelwidth=!, labelindent=0pt, nosep]

\item {\em Time overhead per request:} The time required to receive, process, and forward a task by the middlebox. 

\item {\em Percent of reuse:} The percent of offloaded tasks that reused the results of previously executed tasks.


\item {\em Percent of distributed tasks per edge server:} The number of tasks distributed to each edge server by a middlebox as a percentage of all the tasks offloaded by user devices (clients).

\item {\em Accuracy of reuse:} The percentage of reused tasks, where reuse was accurate. The reuse of a previous task to satisfy an incoming task is accurate if the results of the reused task and the results of the incoming task would have been the same, if the incoming task had been executed from scratch.



\end{enumerate}

\noindent\textbf{Approaches for comparison:} 
First, we compare different approaches that can be used to facilitate computation reuse as part of the \sol design. These approaches include:


\begin{enumerate} [wide, labelwidth=!, labelindent=0pt, nosep]

\item {\em Reuse ideal:} This is the ideal case for reuse, represented as if a single edge server exists that has adequate resources to execute all offloaded tasks. This approach results in the maximum percentage of reuse possible for each dataset.


\item {\em Reuse vanilla:} This is a static approach, which distributes the hash value space equally among edge servers and this distribution does not change afterwards. This approach is equivalent to replacing consistent hashing with LSH and applying the hash function to task input data instead of headers.


\item {\em Reuse mini-buckets:} For \textit{n} available edge servers, the whole hash value space is first divided into \textit{n} slices. Each slice is further divided into \textit{n} smaller slices, and each smaller slice is assigned to a server. This approach is also static in the sense that the size of each smaller slice and the assignment of smaller slices to edge servers do not change over time.

\item {\em Reuse adaptive:} This is a hybrid approach where the hash value space is initially distributed equally among edge servers. After this initial equal distribution, this approach adapts to the conditions (load) at the edge by adjusting the granularity of slices based on Equations~\ref{eq1} and~\ref{eq2} presented in Section~\ref{subsec:prototype}.


  
\end{enumerate}


Second, we compare the \sol to off-the-shelf load balancing approaches that Nginx offers (\eg round robin, random, least connection, consistent hashing). 

\subsection{Evaluation Results}
\label{subsec:results}

In this section, we present results on the trade-offs of the \sol and we compare it to different load balancing approaches. Due to space limitations, we present results for three of the used datasets (MNIST, ASL, and Hololens datasets of Table~\ref{tab:datasets}). Nevertheless, we have verified that the results and the trends presented below hold for all datasets of Table~\ref{tab:datasets}.

\subsubsection{\sol Design Tradeoffs}

\noindent \textbf{Time overhead:} For all approaches (reuse ideal, reuse vanilla, reuse mini-buckets, and reuse adaptive), we evaluate the time needed by the \sol to distribute a task for varying sizes of input data (up to 5MB). Our results show that the \sol needs up to 1.76ms to distribute a task, which is roughly the same amount of time as other load balancing approaches (\eg round robin, random, consistent hashing) implemented by Ngnix. Specifically, the time needed for the \sol to distribute a task is 1\% lower to 3\% higher than the time needed for other load balancing approaches. All approaches are able to distribute traffic at line rate (a total of 30Gbps) for various mixes of tasks and input data sizes.

\begin{table}%
\vspace{0.2cm}
\centering
\caption{Percent of reuse for different datasets.}
\vspace{-0.2cm}
\label{Figure:reuse_percent}
\resizebox{0.65\columnwidth}{!}{%
\centering

\begin{tabular}{|c|c|c|c|c|}
\hline
\begin{tabular}[c]{@{}c@{}}Number of servers  \end{tabular}  & Approach & MNIST & \begin{tabular}[c]{@{}c@{}}ASL
\end{tabular} & \begin{tabular}[c]{@{}c@{}}Hololens
\end{tabular}  \\ \hline

\multirow{4}{*}{3} & Reuse ideal & 18.61 & 51.84 & 61.53 \\ \cline{2-5} 
& Reuse vanilla & 15.47 & 46.73 & 61.48 \\ \cline{2-5} 
& Reuse mini-buckets & 14.65 & 43.50 & 58.50 \\ \cline{2-5}
& Reuse adaptive & 11.39 & 41.13 & 55.65 \\ \hline

\multirow{4}{*}{6} & Reuse ideal & 18.61 & 51.84 & 61.53 \\ \cline{2-5} 
& Reuse vanilla & 13.96 & 43.40 & 56.11 \\ \cline{2-5} 
& Reuse mini-buckets & 11.76 & 41.40 & 53.40 \\ \cline{2-5}
& Reuse adaptive & 10.39 & 39.91 & 51.51 \\ \hline

\multirow{4}{*}{9} & Reuse ideal & 18.61 & 51.84 & 61.53 \\ \cline{2-5} 
& Reuse vanilla & 12.65 & 42.14 & 53.12 \\ \cline{2-5} 
& Reuse mini-buckets & 10.45 & 41.82 & 50.22 \\ \cline{2-5}
& Reuse adaptive & 9.71 & 39.87 & 49.89 \\ \hline

\end{tabular}
}

\vspace{-0.7cm}
\end{table}

\begin{figure*}[htb!]
\captionsetup[subfigure]{aboveskip=-0.00000000000000001pt,belowskip=-0.00000000000000001pt}
	\centering
	\begin{subfigure}[b]{.32\textwidth}
		\centering
		\includegraphics[scale=0.17]{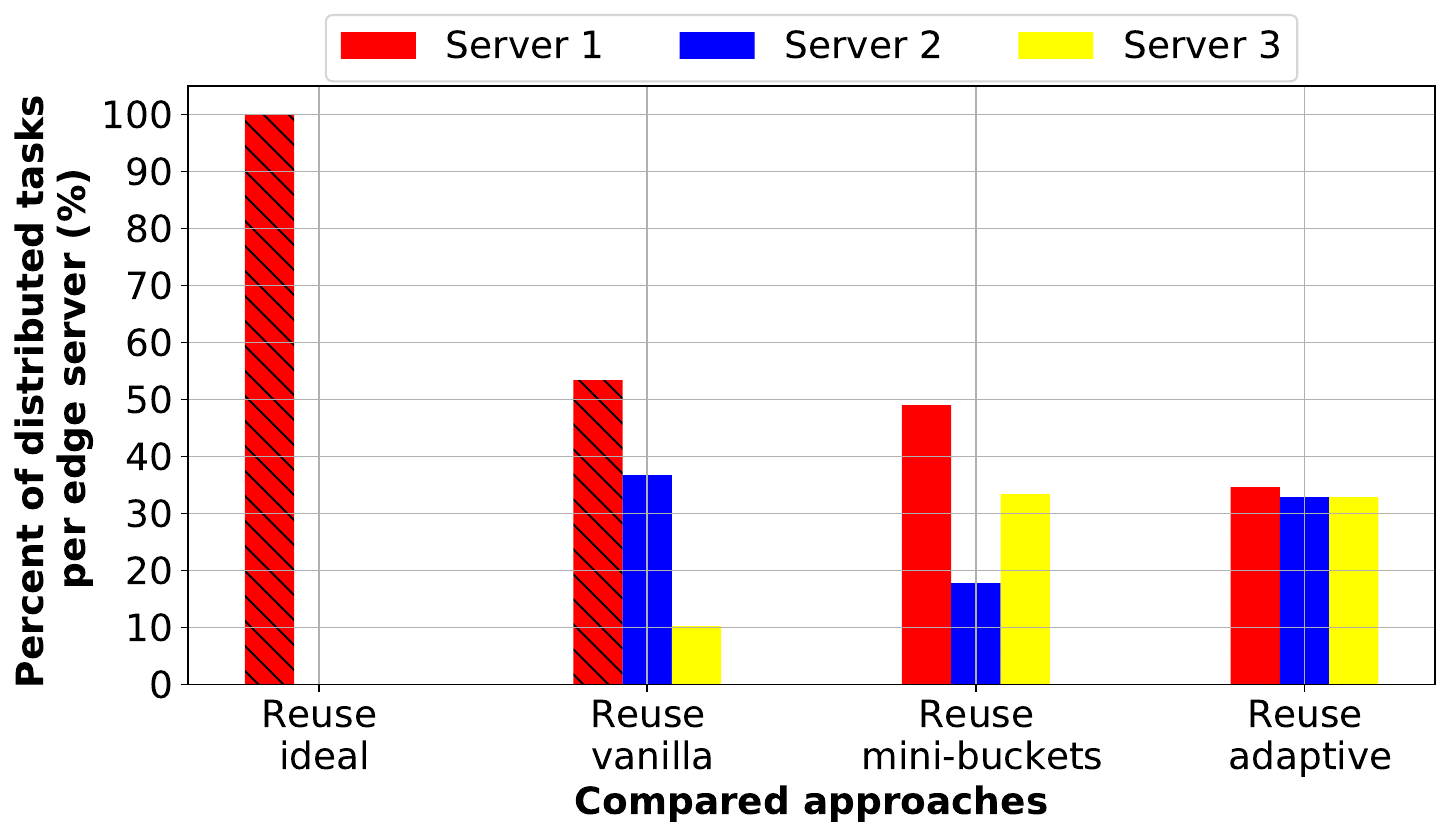}
		\caption{3 edge servers}
		\label{Figure:load_mnist_3}
	\end{subfigure} \hfil
	\begin{subfigure}[b]{.32\textwidth}
	    \centering
	    \includegraphics[scale=0.17]{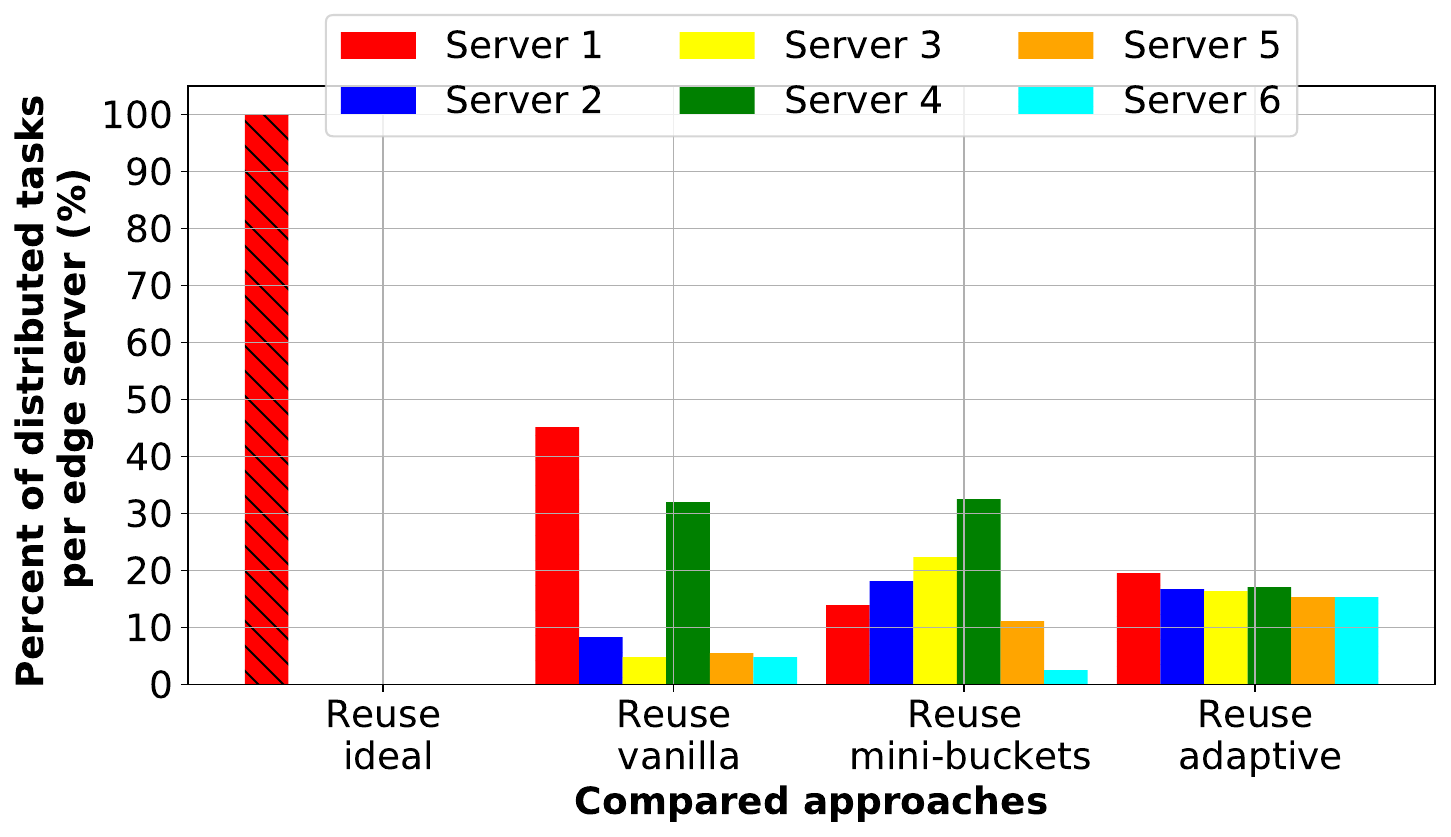}
	    \caption{6 edge servers}
	    \label{Figure:load_mnist_6}
	\end{subfigure}\hfil
	\begin{subfigure}[b]{.32\textwidth}
	    \centering
	    \includegraphics[scale=0.17]{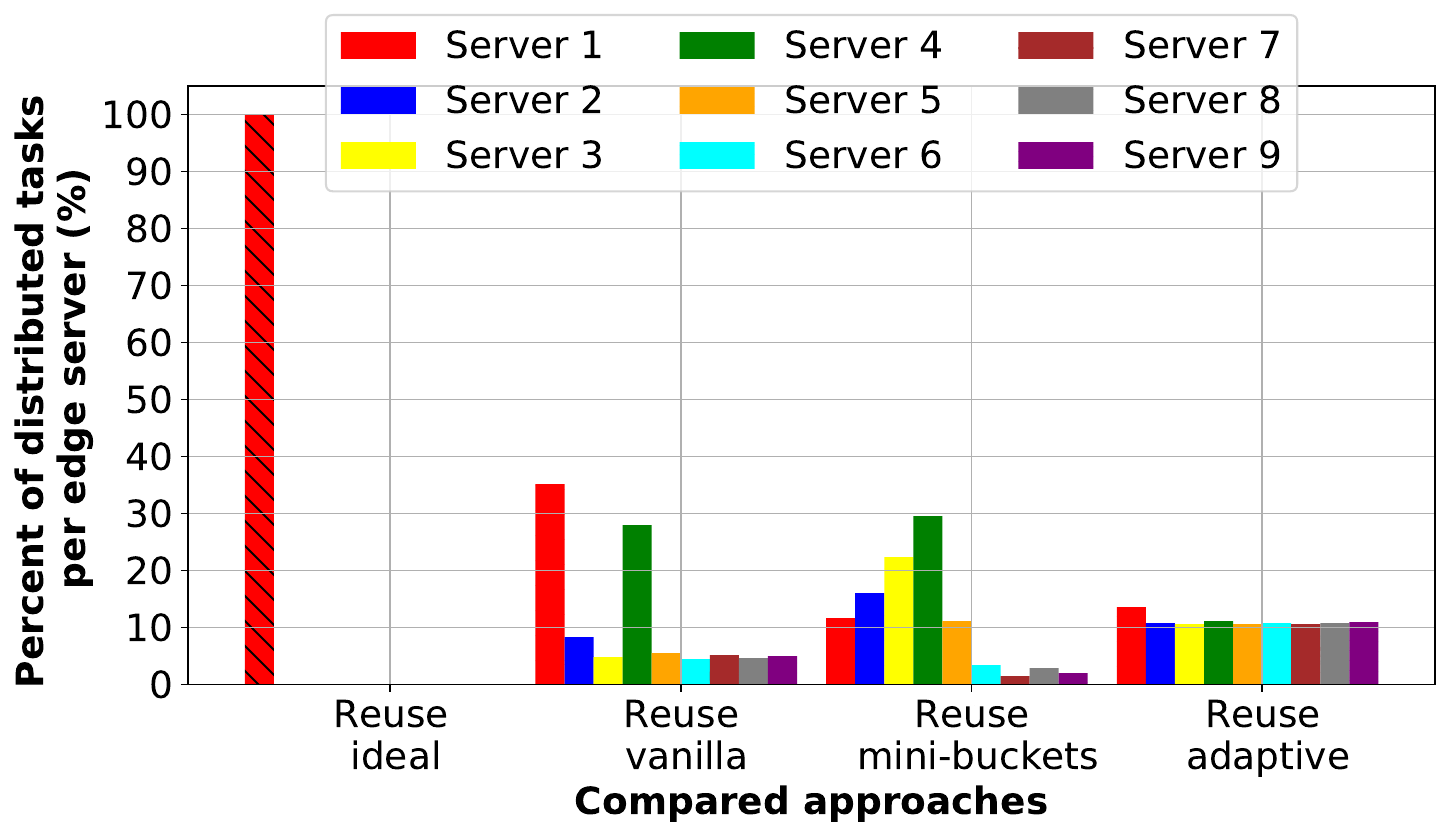}
	    \caption{9 edge servers}
	    \label{Figure:load_mnist_9}
	\end{subfigure}\hfil
	\vspace{-0.25cm}
	\caption{\label{Figure:load_mnist} Load distribution among edge servers for the MNIST dataset.}
	\vspace{-0.5cm}
\end{figure*}

\begin{figure*}[htb!]
\captionsetup[subfigure]{aboveskip=-0.00000000000000001pt,belowskip=-0.00000000000000001pt}
	\centering
	\begin{subfigure}[b]{.32\textwidth}
		\centering
		\includegraphics[scale=0.18]{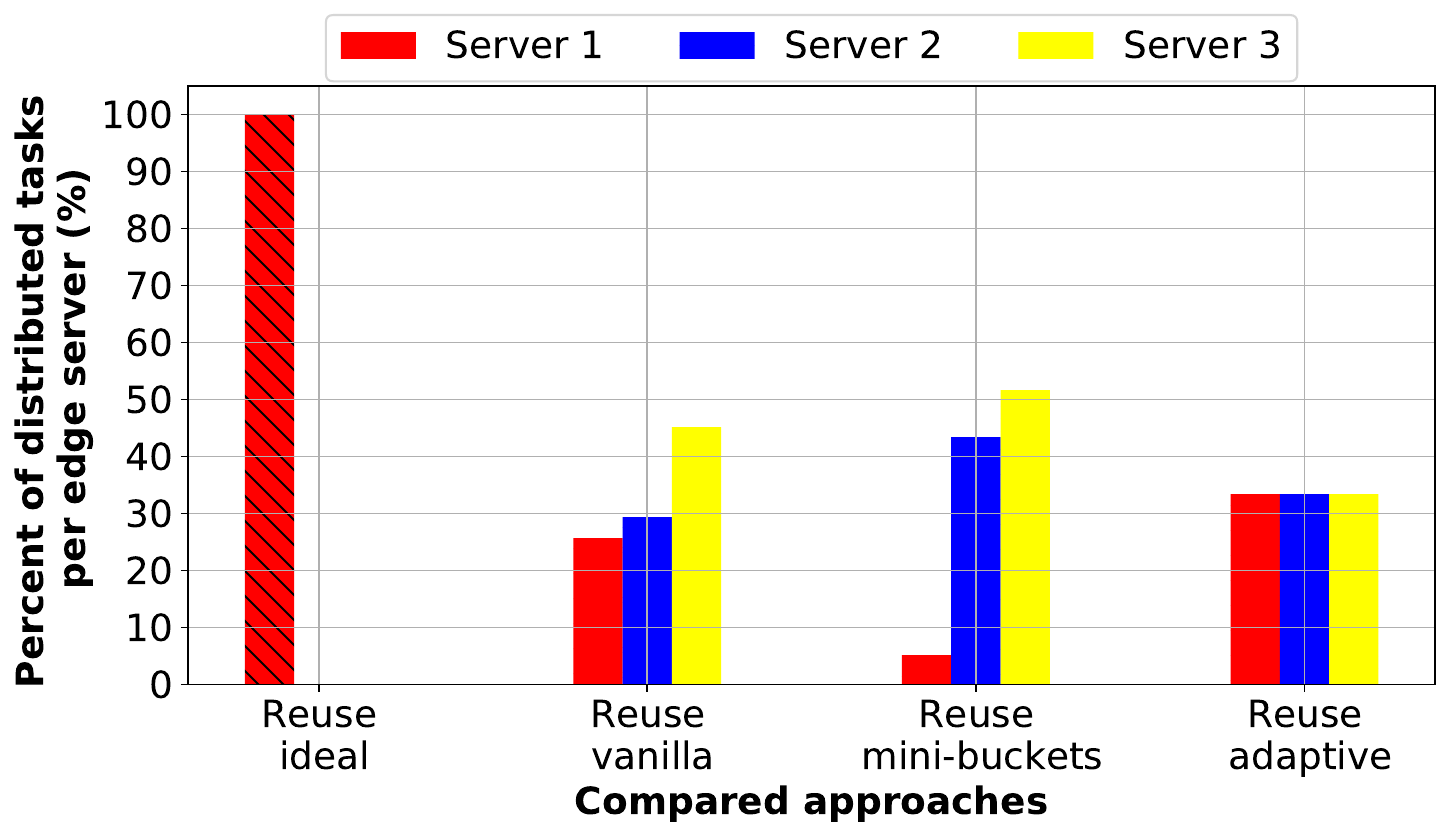}
		\caption{3 edge servers}
		\label{Figure:load_asl_3}
	\end{subfigure} \hfil
	\begin{subfigure}[b]{.32\textwidth}
	    \centering
	    \includegraphics[scale=0.18]{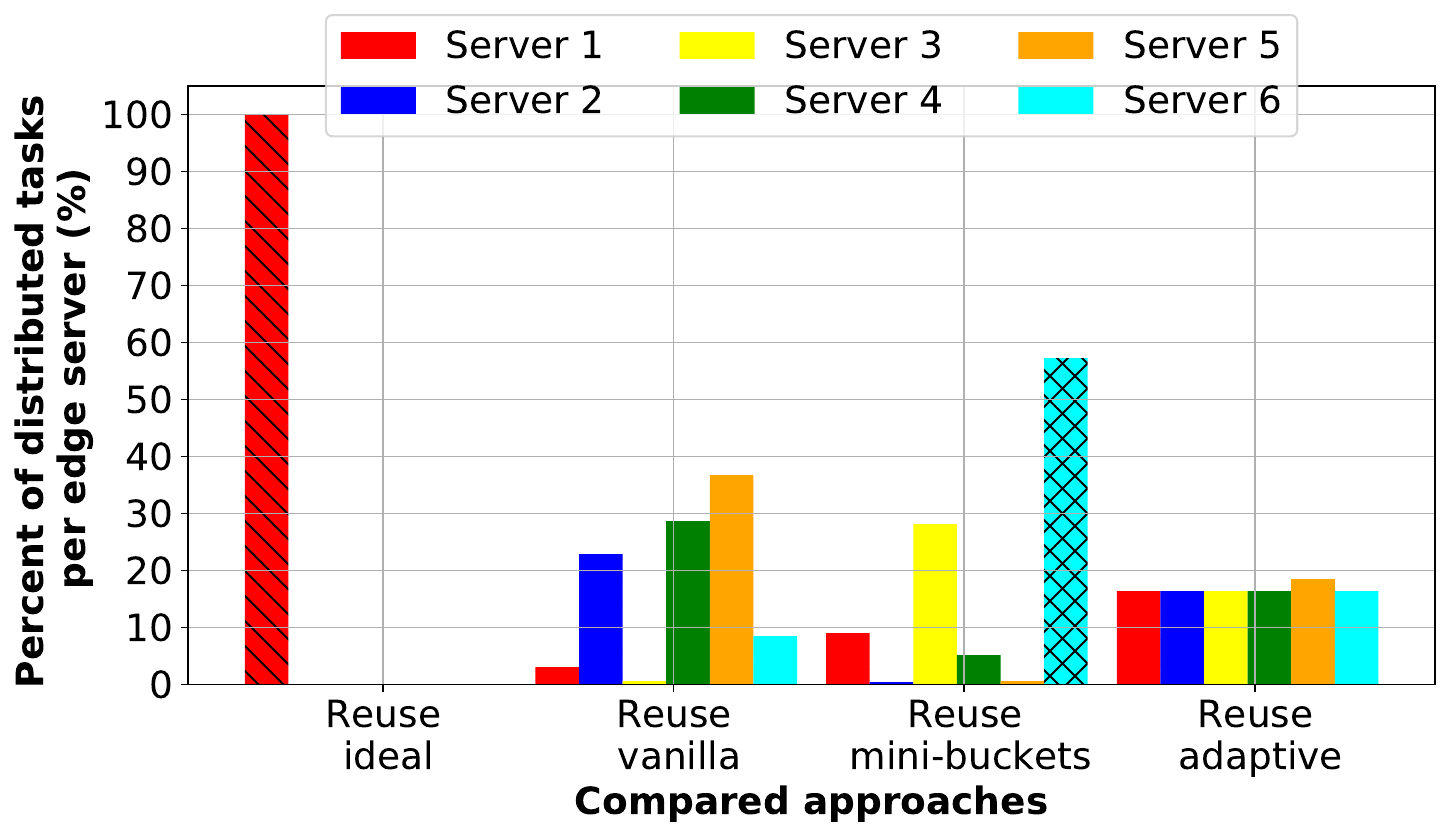}
	    \caption{6 edge servers}
	    \label{Figure:load_asl_6}
	\end{subfigure}\hfil
	\begin{subfigure}[b]{.32\textwidth}
	    \centering
	    \includegraphics[scale=0.18]{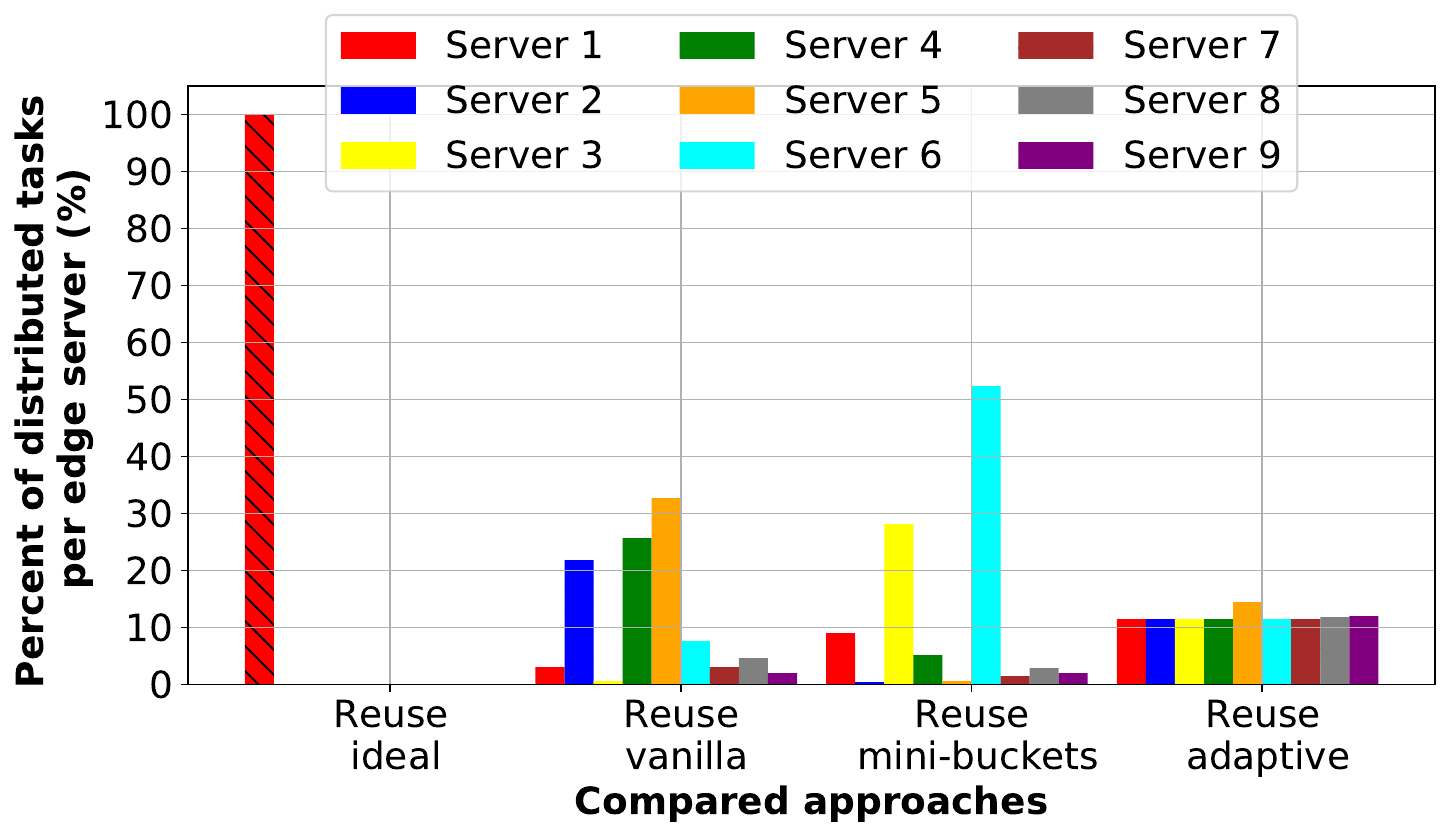}
	    \caption{9 edge servers}
	    \label{Figure:load_asl_9}
	\end{subfigure}\hfil
	\vspace{-0.25cm}
	\caption{\label{Figure:load_asl} Load distribution among edge servers for the ASL dataset.}
	\vspace{-0.5cm}
\end{figure*}

\begin{figure*}[htb!]
\captionsetup[subfigure]{aboveskip=-0.00000000000000001pt,belowskip=-0.00000000000000001pt}
	\centering
	\begin{subfigure}[b]{.32\textwidth}
		\centering
		\includegraphics[scale=0.18]{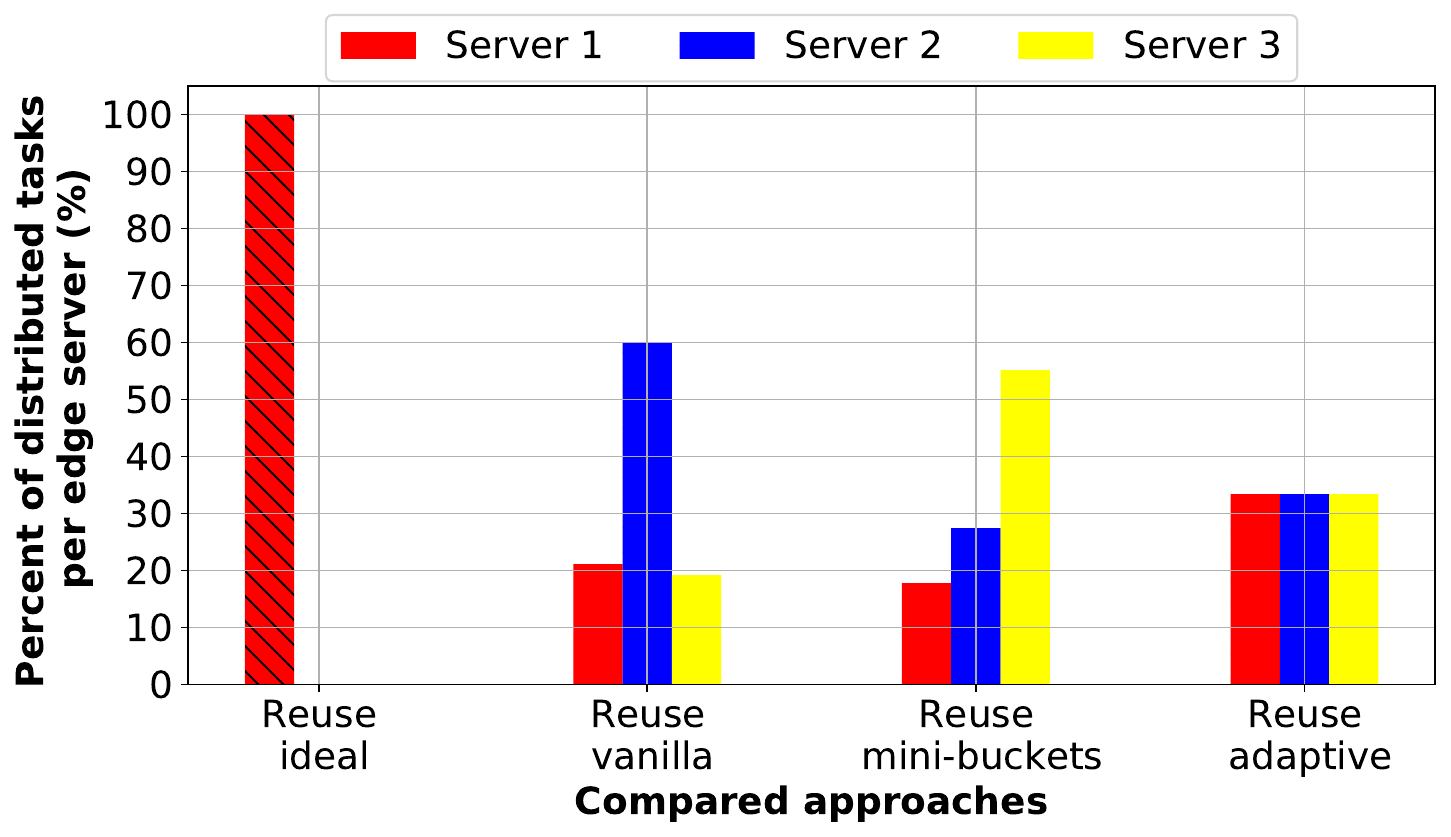}
		\caption{3 edge servers}
		\label{Figure:load_hololens_3}
	\end{subfigure} \hfil
	\begin{subfigure}[b]{.32\textwidth}
	    \centering
	    \includegraphics[scale=0.18]{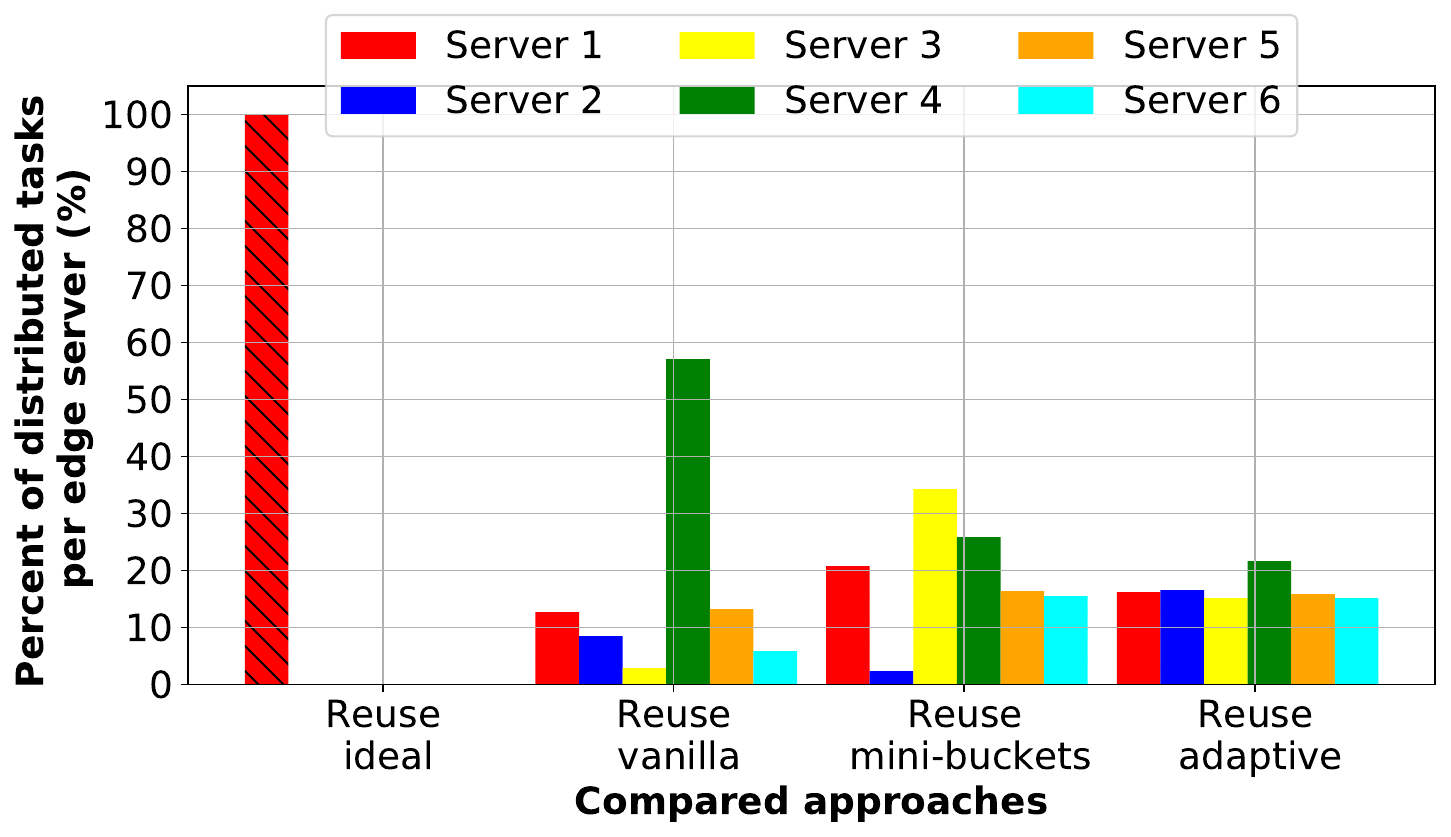}
	    \caption{6 edge servers}
	    \label{Figure:load_hololens_6}
	\end{subfigure}\hfil
	\begin{subfigure}[b]{.32\textwidth}
	    \centering
	    \includegraphics[scale=0.18]{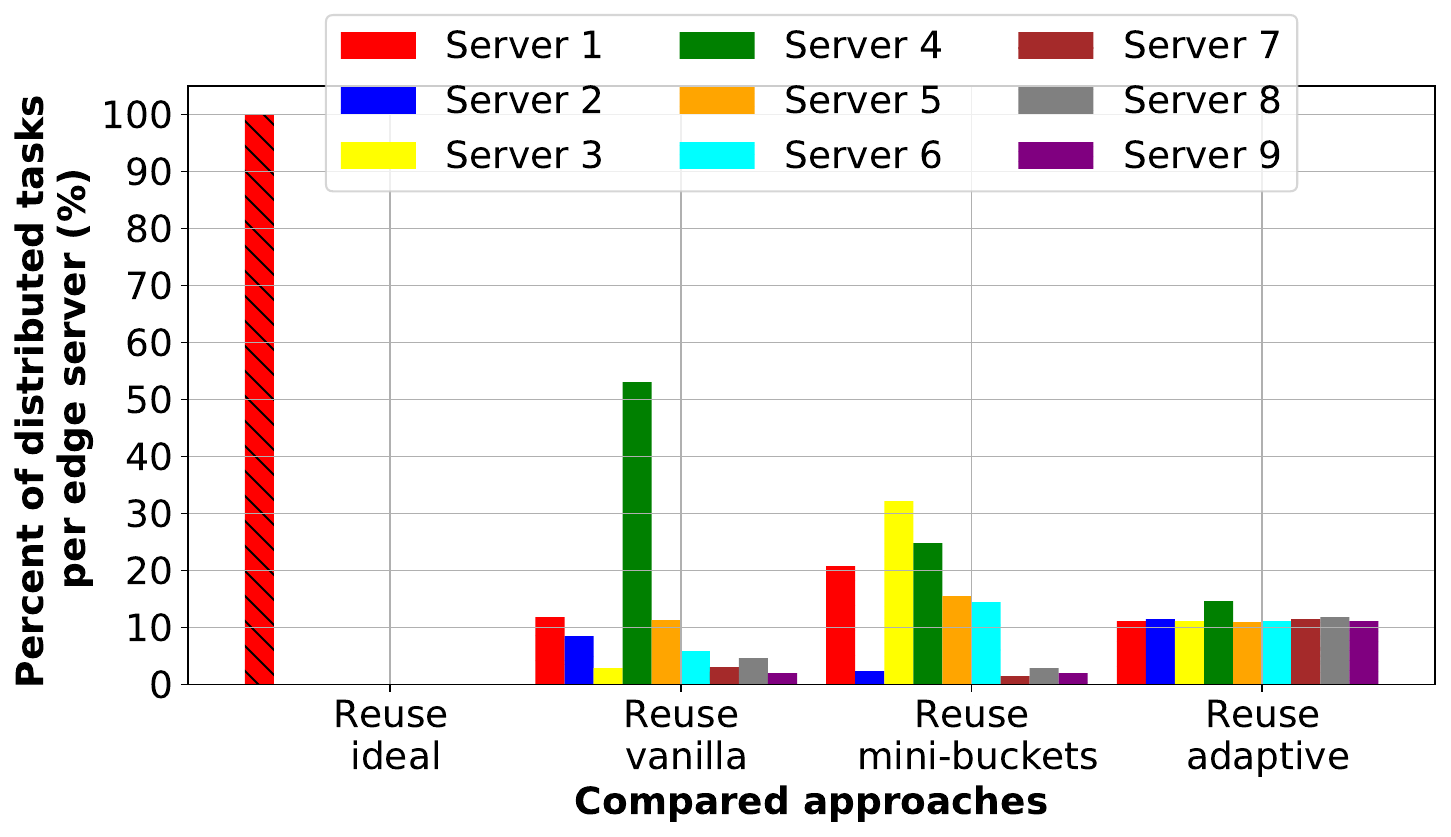}
	    \caption{9 edge servers}
	    \label{Figure:load_hololens_9}
	\end{subfigure}\hfil
	\vspace{-0.25cm}
	\caption{\label{Figure:load_holo} Load distribution among edge servers for the Hololens dataset.}
	\vspace{-0.4cm}
\end{figure*}

\noindent \textbf{Percent of reuse:} 
As shown in Table~\ref{Figure:reuse_percent}, reuse decreases up to 12\% in the case of the reuse adaptive approach compared to the ideal reuse case. 
Compared to static approaches (reuse vanilla and reuse mini-buckets), reuse adaptive results in a reuse reduction of less than 5\%. This reduction is due to the redistribution of the hash value space performed by reuse adaptive to balance task distribution among servers. 

\noindent \textbf{Percent of distributed tasks per edge server:} 
Our results (Figures~\ref{Figure:load_mnist},~\ref{Figure:load_asl}, and~\ref{Figure:load_holo}) show that the reuse adaptive approach distributes tasks substantially more uniformly among servers compared to other approaches. Reuse adaptive also redistributes the hash value space among servers, so that load imbalances are identified and mitigated. Static approaches (reuse vanilla and reuse mini-buckets) are less effective, since they do not redistribute the hash value space in cases of load imbalances. Finally, the reuse ideal approach distributes all tasks to a single server, thus imposing all the load to this server (\eg server 1 in the results of Figures~\ref{Figure:load_mnist},~\ref{Figure:load_asl}, and~\ref{Figure:load_holo}).

\noindent \textbf{Accuracy of reuse:} For all approaches, the \sol facilitates the reuse of computation with accuracy that exceeds 90\% across all datasets (in some cases even reaching 100\%). 

\noindent \textbf{Conclusion:} The reuse adaptive approach achieves the best trade-off between reuse and load balancing. Reuse adaptive can effectively balance the distribution of tasks among edge servers with a minor reduction of the reuse percentage (less than 5\%) compared to static approaches. To this end, for the results presented in Section~\ref{subsubsec:comp}, we implement the \sol with the reuse adaptive approach.


\subsubsection{Comparison to Different Load Balancing Approaches}
\label{subsubsec:comp}

\noindent \textbf{Time overhead:} We evaluate the time needed by the \sol and other load balancing approaches to distribute a task for varying sizes of input data (up to 5MB). 
Our results show that the \sol requires
roughly the same amount of time to distribute a task as other load balancing approaches (up to 1.76ms per task, which is 1\% lower to 3\% higher than the time needed by other load balancing approaches). We experimented with different traffic mixes that consist of tasks with various input sizes and we verified that the \sol is able to distribute traffic at line rate (a total of 30Gbps). 

\begin{table}[htb!]
\centering
\vspace{-0.3cm}
\caption{Percent of reuse for different datasets.} 
\vspace{-0.2cm}
\resizebox{0.6\columnwidth}{!}{%
\label{Figure:compare_reuse_percent}
\centering

\begin{tabular}{|c|c|c|c|c|}
\hline
\begin{tabular}[c]{@{}c@{}}Number of servers  \end{tabular}  & Approach & MNIST & \begin{tabular}[c]{@{}c@{}}ASL
\end{tabular} & \begin{tabular}[c]{@{}c@{}}Hololens
\end{tabular}  \\ \hline

\multirow{5}{*}{3} & Round Robin & 9.95 & 38.28 & 43.89 \\ \cline{2-5} 
& Random & 10.05 & 37.89 & 43.55 \\ \cline{2-5} 
& Least connection & 10.67 & 37.87 & 43.23 \\ \cline{2-5}
& Consistent hashing & 9.69 & 38.06 & 43.98 \\ \cline{2-5}
& Reuse adaptive & 11.39 & 41.13 & 55.65 \\ \hline

\multirow{5}{*}{6} & Round Robin & 6.03 & 23.76 & 32.59 \\ \cline{2-5} 
& Random & 6.01 & 23.09 & 32.62 \\ \cline{2-5} 
& Least connection & 6.51 & 23.34 & 32.87 \\ \cline{2-5}
& Consistent hashing & 6.80 & 23.54 & 32.76 \\ \cline{2-5}
& Reuse adaptive & 10.03 & 38.77 & 51.56 \\ \hline

\multirow{5}{*}{9} & Round Robin & 4.04 & 20.67 & 26.78 \\ \cline{2-5} 
& Random & 4.01 & 20.11 & 26.82 \\ \cline{2-5} 
& Least connection & 4.50 & 20.14 & 26.91 \\ \cline{2-5}
& Consistent hashing & 4.76 & 20.23 & 26.87 \\ \cline{2-5}
& Reuse adaptive & 9.71 & 35.87 & 46.94 \\ \hline

\end{tabular}
}
\vspace{-0.3cm}
\end{table}

\begin{figure*}[htb!]
\vspace{-0.1cm}
\captionsetup[subfigure]{aboveskip=-0.00000000000000001pt,belowskip=-0.00000000000000001pt}
	\centering
	\begin{subfigure}[b]{.32\textwidth}
		\centering
		\includegraphics[scale=0.17]{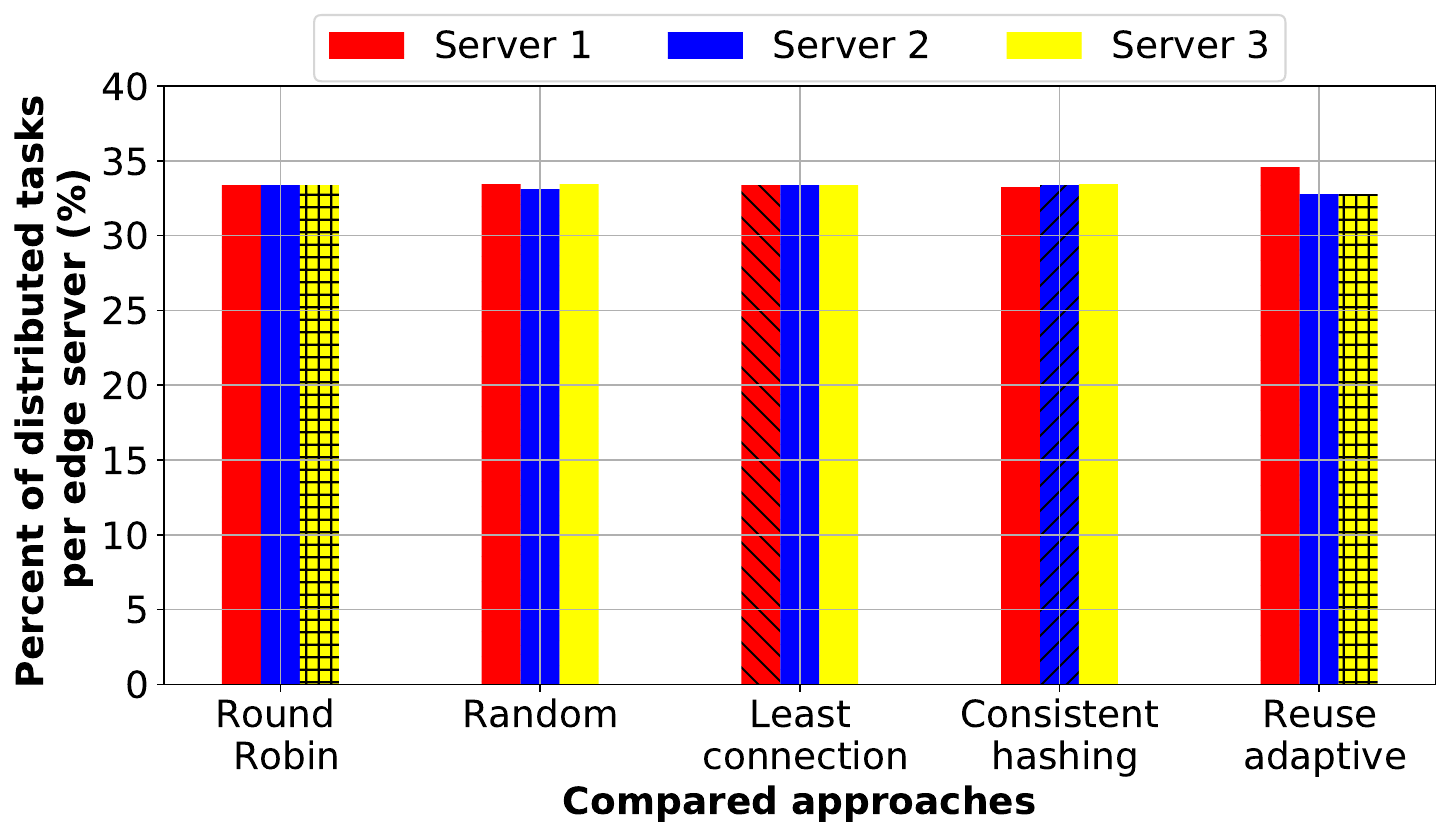}
		\caption{3 edge servers}
		\label{Figure:compare_load_mnist_3}
	\end{subfigure} 
	\begin{subfigure}[b]{.32\textwidth}
	    \centering
	    \includegraphics[scale=0.17]{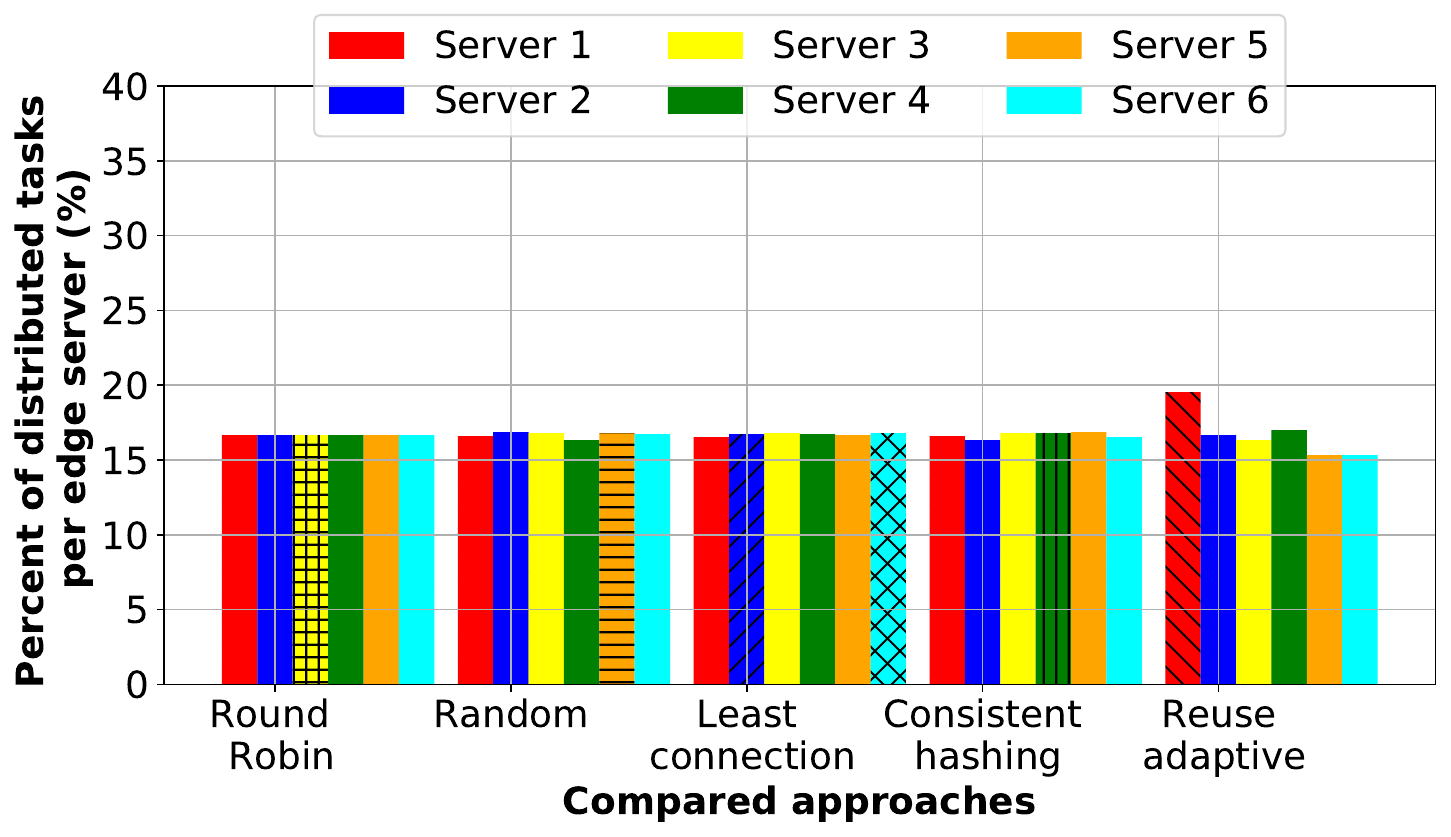}
	    \caption{6 edge servers}
	    \label{Figure:compare_load_mnist_6}
	\end{subfigure}
	\begin{subfigure}[b]{.32\textwidth}
	    \centering
	    \includegraphics[scale=0.17]{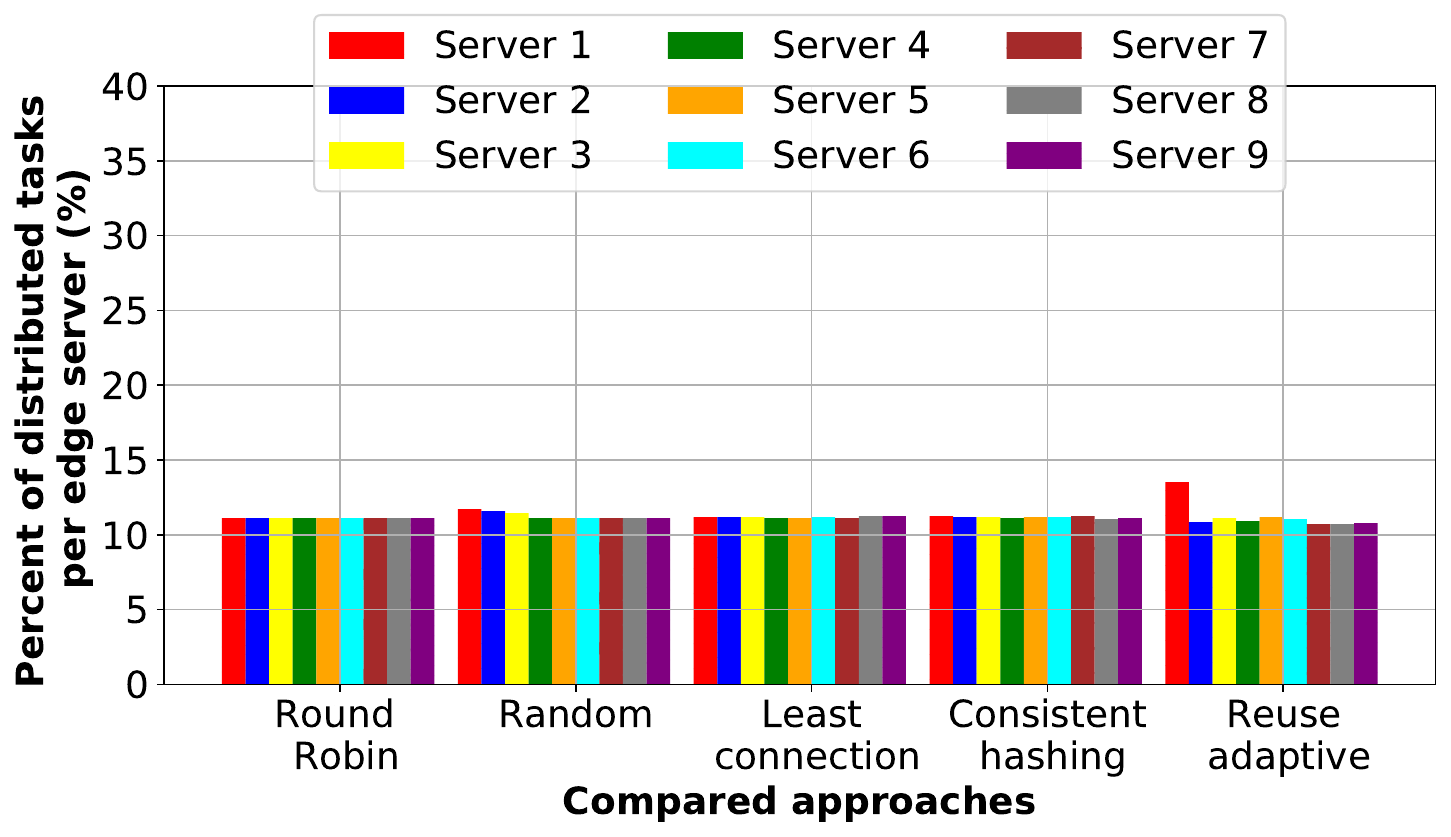}
	    \caption{9 edge servers}
	    \label{Figure:compare_load_mnist_9}
	\end{subfigure}
	\vspace{-0.25cm}
	\caption{\label{Figure:load_mnist_comp} Load distribution among edge servers for the MNIST dataset.}
	\vspace{-0.45cm}
\end{figure*}

\begin{figure*}[h!]
\captionsetup[subfigure]{aboveskip=-0.00000000000000001pt,belowskip=-0.00000000000000001pt}
	\centering
	\begin{subfigure}[b]{.32\textwidth}
		\centering
		\includegraphics[scale=0.17]{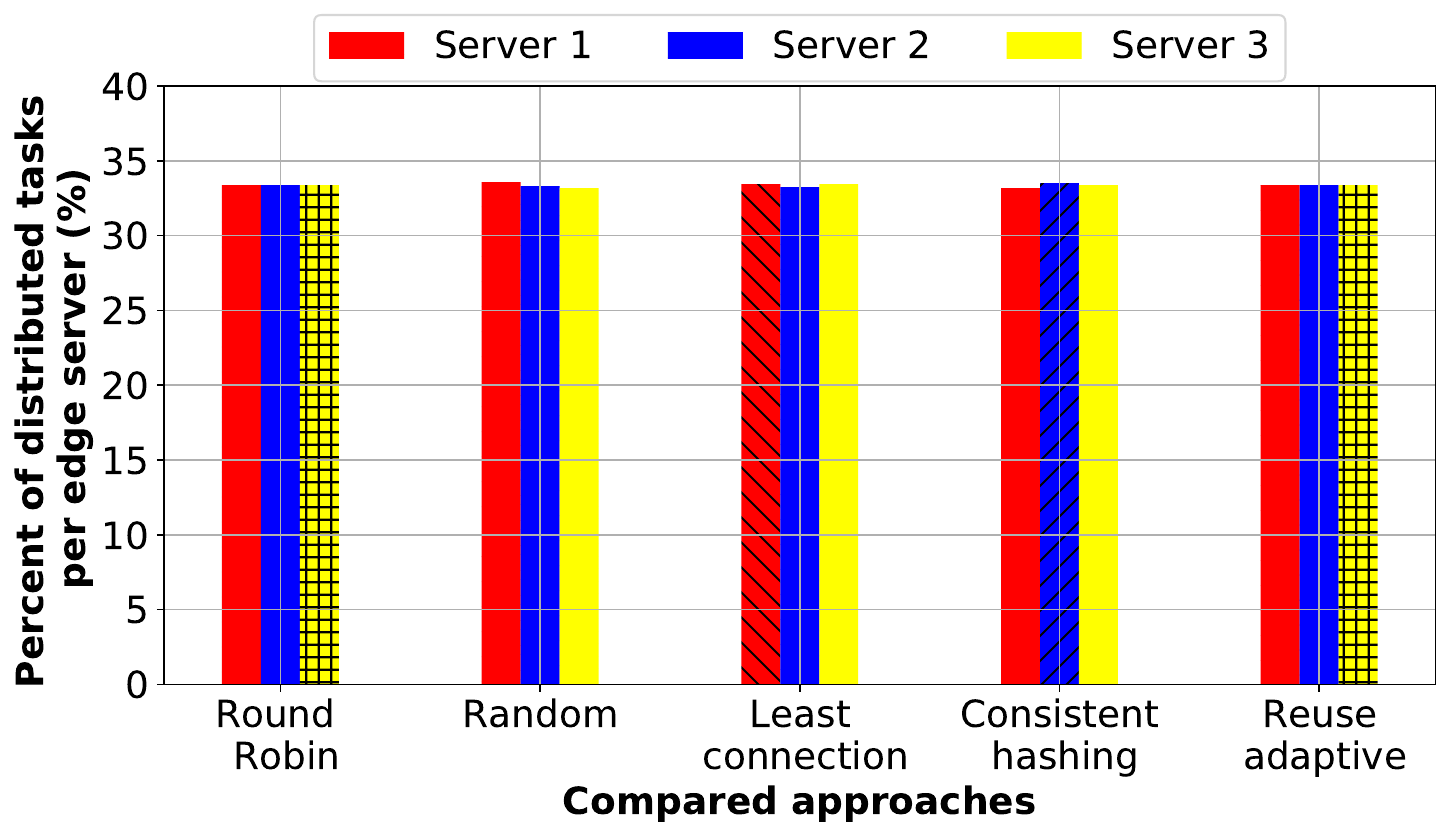}
		\caption{3 edge servers}
		\label{Figure:compare_load_asl_3}
	\end{subfigure} \hfil
	\begin{subfigure}[b]{.32\textwidth}
	    \centering
	    \includegraphics[scale=0.17]{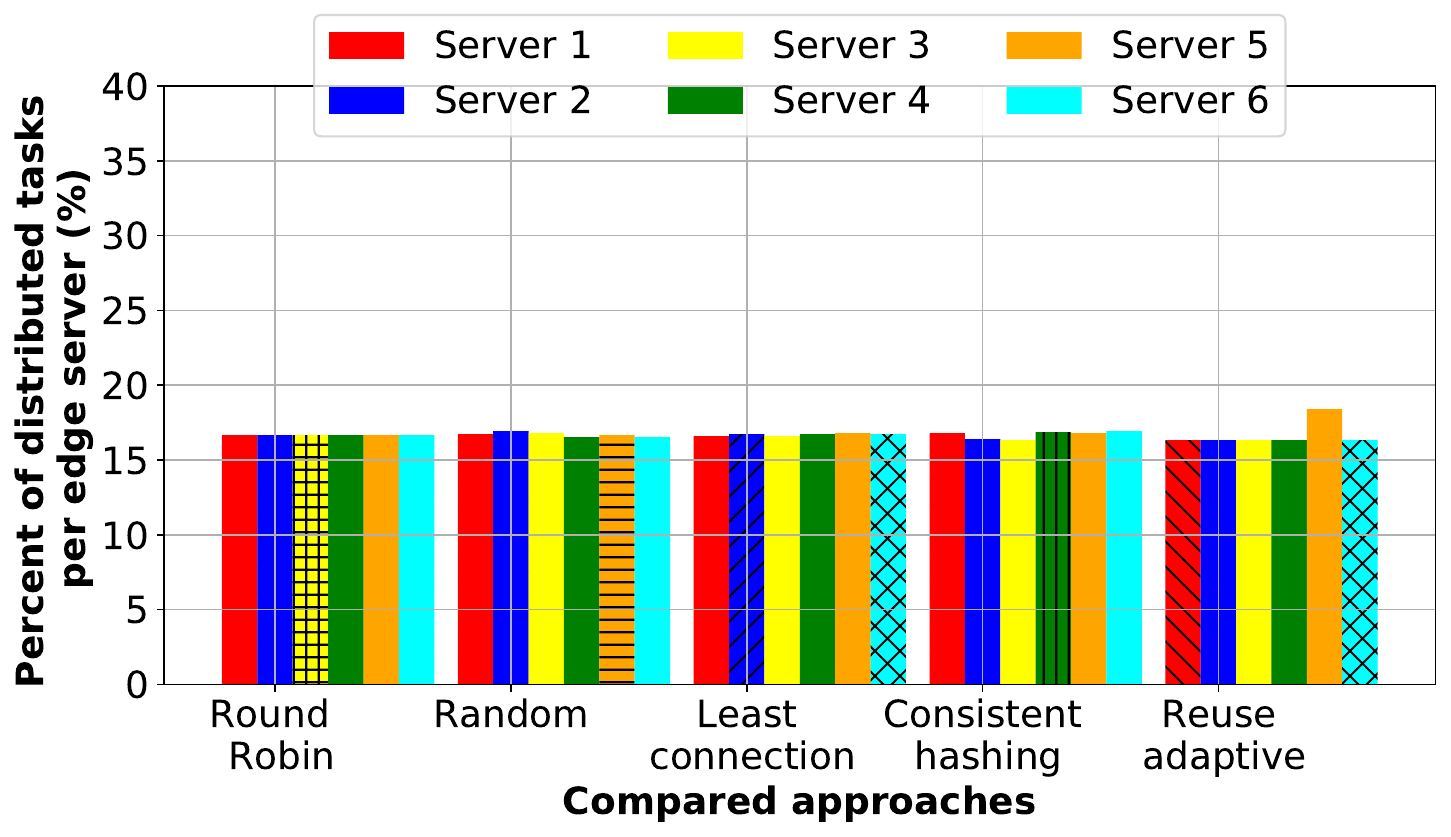}
	    \caption{6 edge servers}
	    \label{Figure:compare_load_asl_6}
	\end{subfigure}\hfil
	\begin{subfigure}[b]{.32\textwidth}
	    \centering
	    \includegraphics[scale=0.17]{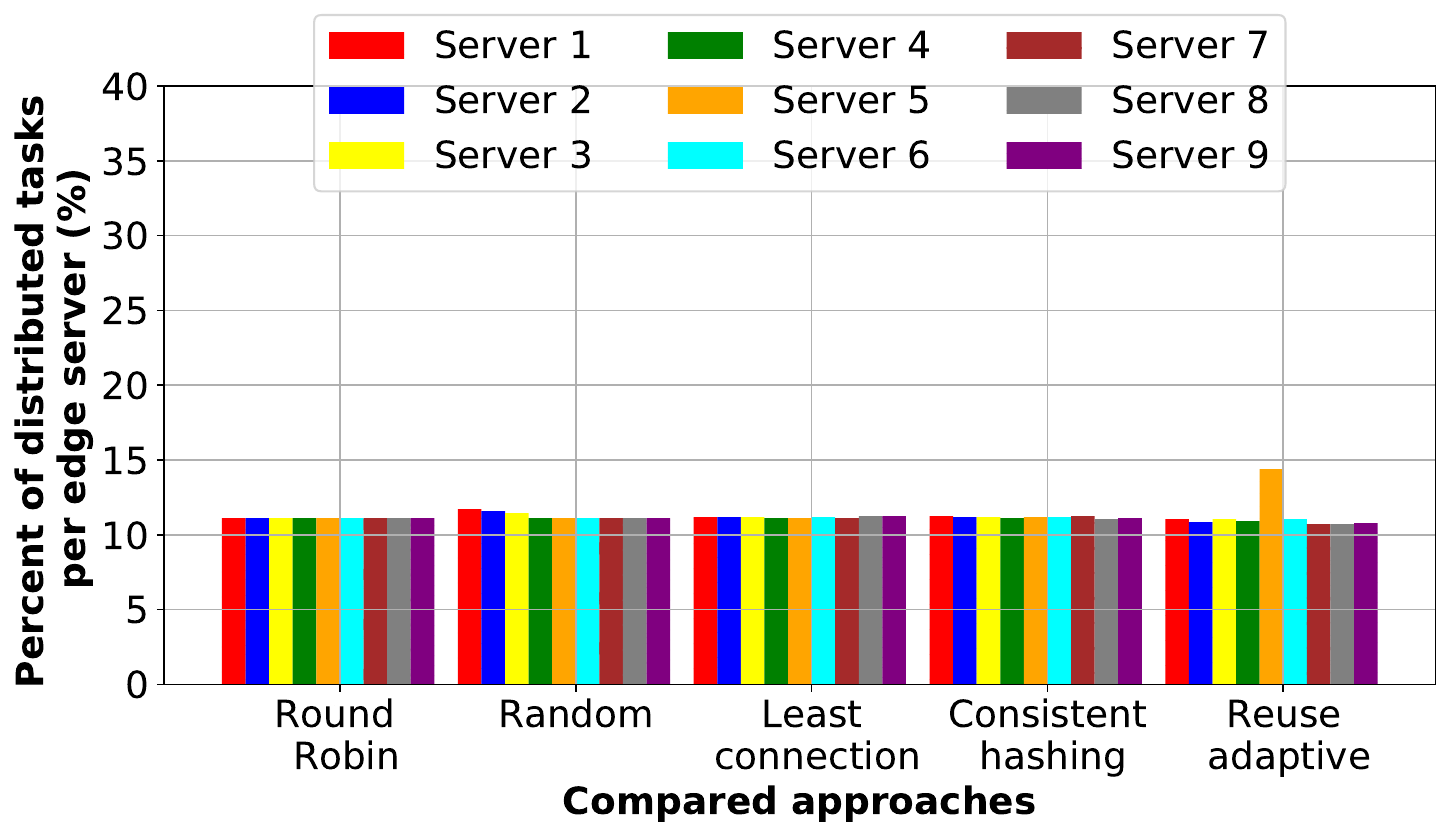}
	    \caption{9 edge servers}
	    \label{Figure:compare_load_asl_9}
	\end{subfigure}\hfil
	\vspace{-0.25cm}
	\caption{\label{Figure:load_asl_comp} Load distribution among edge servers for the ASL dataset.}
	\vspace{-0.45cm}
\end{figure*}

\begin{figure*}[h!]
\captionsetup[subfigure]{aboveskip=-0.00000000000000001pt,belowskip=-0.00000000000000001pt}
	\centering
	\begin{subfigure}[b]{.32\textwidth}
		\centering
		\includegraphics[scale=0.17]{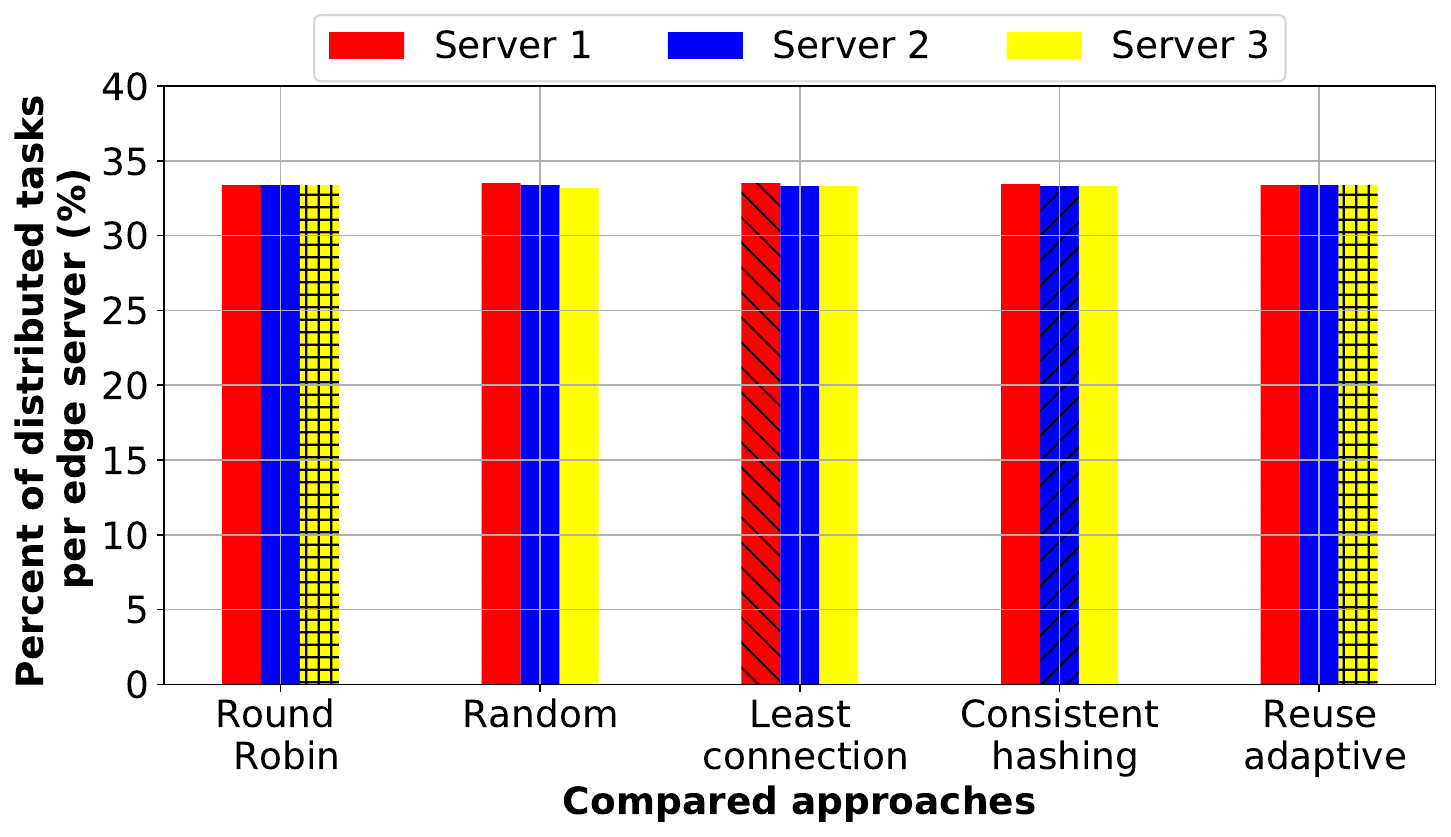}
		\caption{3 edge servers}
		\label{Figure:compare_load_hololens_3}
	\end{subfigure} \hfil
	\begin{subfigure}[b]{.32\textwidth}
	    \centering
	    \includegraphics[scale=0.17]{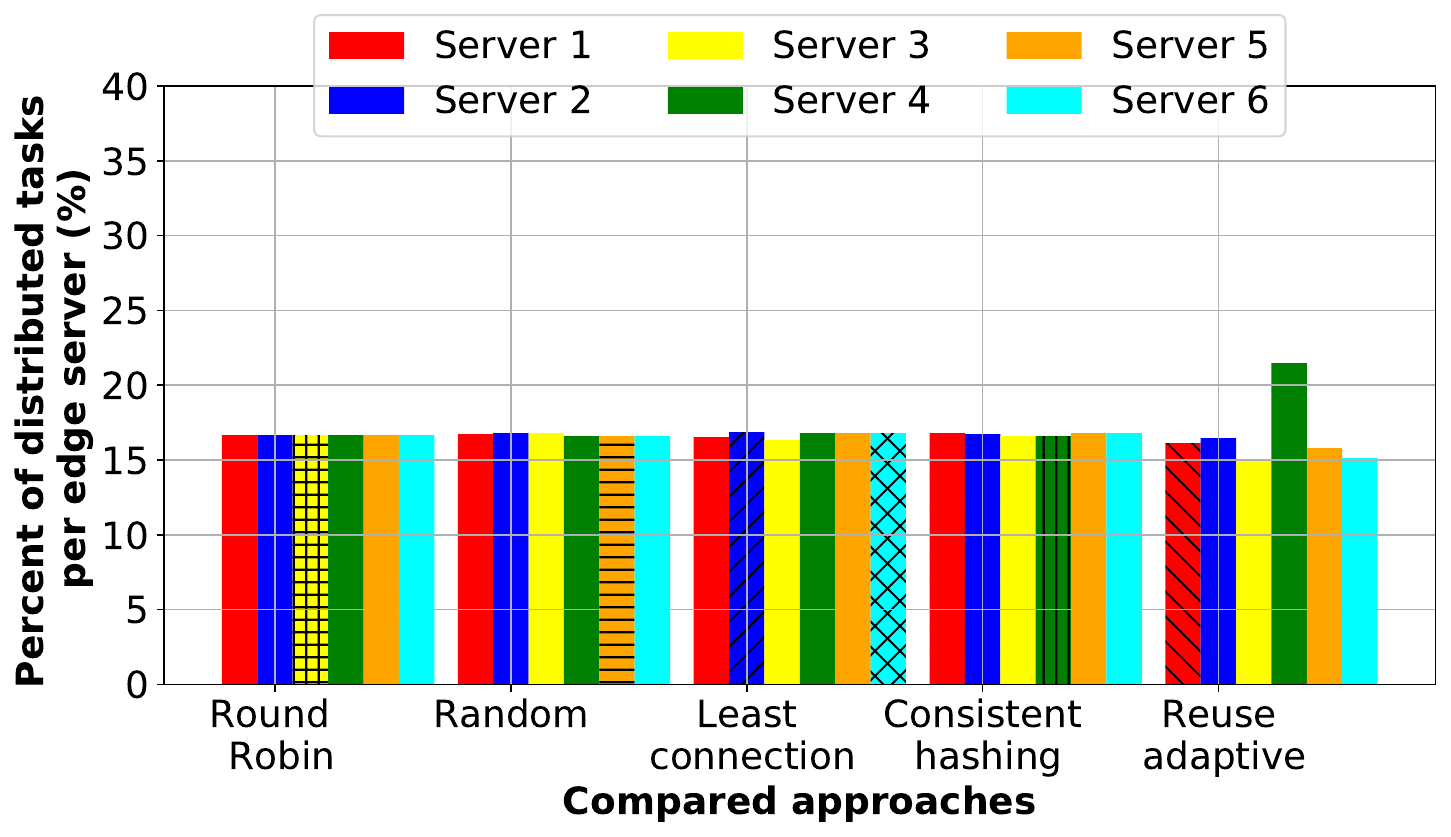}
	    \caption{6 edge servers}
	    \label{Figure:compare_load_hololens_6}
	\end{subfigure}\hfil
	\begin{subfigure}[b]{.32\textwidth}
	    \centering
	    \includegraphics[scale=0.17]{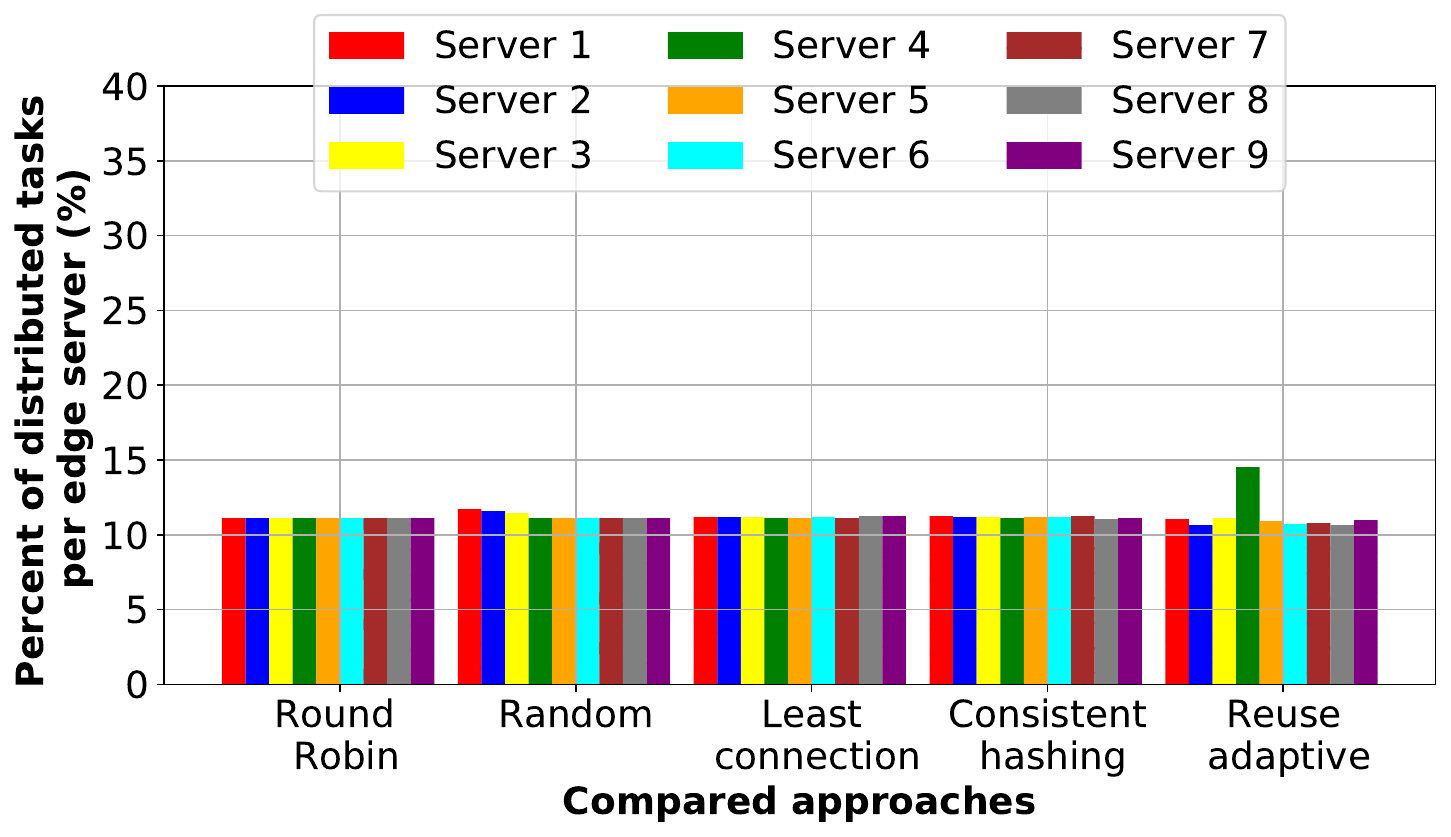}
	    \caption{9 edge servers}
	    \label{Figure:compare_load_hololens_9}
	\end{subfigure}\hfil
	\vspace{-0.25cm}
	\caption{\label{Figure:load_holo_comp} Load distribution among edge servers for the Hololens dataset.}
	\vspace{-0.6cm}
\end{figure*}

\noindent \textbf{Percent of reuse:} 
Our results in Table~\ref{Figure:compare_reuse_percent} 
show that the \sol achieves up to 20\% higher percent of reuse than other load balancing approaches, since the \sol is aware of the semantics of reuse, thus being able to identify and distribute similar tasks to the same edge server(s). Our results also 
show that the percent of reuse severely degrades for load balancing approaches that are not aware of the reuse semantics 
as we increase the number of servers. This degradation for the \sol is much smaller as we increase the number of servers, since it can identify similar tasks independent of the number of available servers.

\noindent \textbf{Percent of distributed tasks per edge server:} 
Figures~\ref{Figure:load_mnist_comp},~\ref{Figure:load_asl_comp}, and~\ref{Figure:load_holo_comp} show that the \sol balances task distribution among 
servers with minor deviations (less than 5\%) compared to other load balancing approaches, thus mitigating load imbalances that occur due to reuse. These deviations are primarily due to the fact that in datasets with high correlations, such as the Hololens dataset, higher amounts of data typically correspond to only a few hash values in the overall hash value space. In such cases, the \sol creates over time slices of finer granularity (as described in Section~\ref{subsec:slicing}) to balance task distribution among servers. Nevertheless, the fact that different amounts of data are hashed to individual hash values over time results in these minor deviations of task distribution.

\noindent \textbf{Accuracy of reuse:} 
Our results show that the \sol achieves 3-14\% higher reuse accuracy than the compared approaches. The \sol distributes tasks to edge servers that are likely to reuse previously executed tasks, while other approaches are not aware of the reuse semantics. In addition, the \sol consistently achieves reuse accuracy that exceeds 90\%, reaching for certain datasets almost 100\%.

\noindent \textbf{Conclusion:} The \sol increases the percentage of reuse by up to 20\% compared to load balancing approaches that are not aware of the computation reuse semantics. The \sol also balances the distribution of tasks among edge servers with minor deviations (less than 5\%) compared to non computation reuse aware load balancing approaches.

\section{Related Work}
\label{sec:back}

\noindent\textbf{Load balancing:} 
Load balancing can be performed at the transport (layer 4) or the application layer (layer 7). 
Ananta \cite{patel2013ananta} is a layer 4 load balancer that runs on commodity hardware and aims to satisfy the performance and reliability needs of multi-tenant cloud computing environments.
Duet \cite{gandhi2014duet} extended Ananta through the use of existing switch hardware in combination with a small deployment of a software load balancer to achieve low packet processing times. 
RUBIK \cite{gandhi2015rubik} further reduced the bandwidth usage of Duet caused by traffic redirection using locality. Similar to Duet, Silkroad \cite{miao2017silkroad} and Loom \cite{zhang2021loom} use switches as the load balancing instances. 

Yoda \cite{gandhi2016yoda} is a layer 7 load balancer that provides high availability through virtual IP addresses to establish connections with servers and clients. 
Desmouceaux \textit{et al.} proposed a layer 4 load balancing strategy that can distribute the traffic in an application-aware manner similar to layer 7 load balancing but without centralized monitoring~\cite{desmouceaux20186lb}. APRIL~\cite{nadig2019april} is an application-aware load balancer, where a deep recurrent neural network predicts future load information to maximize resource utilization based on application metadata. HAProxy~\cite{haproxy} 
provides application-aware load balancing. None of these approaches, however, has considered incorporating computation reuse semantics into the load balancing process.


\noindent\textbf{Computation reuse:} Cachier~\cite{drolia2017cachier} was among the first proposals to explore a direction, in which edge servers were used as specialized caches to improve the latency for recognition applications by leveraging spatiotemporal locality of requests. 
Guo \textit{et al.} proposed Potluck as a service to deduplicate and reuse the results of tasks with correlated input data across applications running on the same device \cite{guo2018potluck}. 
FoggyCache \cite{guo2018foggycache} was proposed to achieve cross-device computation reuse when nearby devices are running the same application.
Coterie \cite{meng2020coterie} decoupled the near and far background environment frames to enhance the similarity among frames in multiplayer virtual reality and cache the frames to reduce network load. Mastorakis \textit{et al.} proposed ICedge, which facilitates computation reuse in a distributed edge network in a limited fashion~\cite{mastorakis2020icedge}. Al Azad \textit{et al.} discussed the potential of computation reuse and the challenges that arise in distributed edge network settings~\cite{al2022promise}. 
Reservoir \cite{al2022reservoir} was proposed to 
enable reuse at edge devices, edge servers, and within the network infrastructure. None of these approaches, however, has explored the implications of computation reuse on the load distribution among edge servers and the load imbalances that can be caused.

\noindent\textbf{LSH:} Unlike traditional hashing, where the goal is to avoid collisions and generate a unique fixed length hash value for each unique piece of data, LSH is a hashing algorithm that assigns similar inputs to the same buckets with high probability \cite{indyk1998approximate}. LSH can be used to reduce the search space of high dimensional data for nearest neighbor search, where pieces of input data may not be the same but similar to each other. Multiple 
LSH functions can be used simultaneously to improve search accuracy at the cost of maintaining a large number of hash tables \cite{indyk1998approximate}. 
Multi-probe LSH was proposed~\cite{andoni2015practical}
to reduce overhead by searching multiple buckets with similar data for a small number of hash tables.

\section {Conclusion}
\label{sec:conclusion}

In this paper, we presented the \sol, a middlebox that balances the load and facilitates computation reuse in edge computing environments. The \sol design features mechanisms to identify and deduplicate similar 
tasks offloaded by user devices, collect information about the usage of edge servers' resources, manage the addition of new servers and the failures of existing servers, and ultimately balance the load imposed on servers. Our evaluation results showed that the \sol balances the distribution of tasks among edge servers with minor deviations compared to non computation reuse aware load balancing approaches and achieves higher percentages of reuse than these approaches.


\section*{Acknowledgements}

This work is partially supported by the National Science Foundation through awards CNS-2104700 and CNS-2306685, as well as by the U.S. Army Engineer Research \& Development Center (ERDC) through HPC PET special project 5025.

\bibliographystyle{IEEEtran}
\bibliography{refs.bib}
\end{document}